%% file: main.tex
\documentclass[sigconf]{acmart}
\usepackage{multirow}
\usepackage{float}
\usepackage{subfig}
\usepackage{enumitem}
\usepackage{algorithmic}
\usepackage[ruled, linesnumbered]{algorithm2e}

\newcommand{\fullname}{\emph{Graph Bottlenecked Social Recommendation~(GBSR)}}
\newcommand{\shortname}{\emph{GBSR}}

\AtBeginDocument{%
  \providecommand\BibTeX{{%
    \normalfont B\kern-0.5em{\scshape i\kern-0.25em b}\kern-0.8em\TeX}}}

\setcopyright{acmlicensed}

\copyrightyear{2024}
\acmYear{2024}
\setcopyright{acmlicensed}\acmConference[KDD '24]{Proceedings of the 30th ACM SIGKDD Conference on Knowledge Discovery and Data Mining}{August 25--29, 2024}{Barcelona, Spain}
\acmBooktitle{Proceedings of the 30th ACM SIGKDD Conference on Knowledge Discovery and Data Mining (KDD '24), August 25--29, 2024, Barcelona, Spain}
\acmDOI{10.1145/3637528.3671807}
\acmISBN{979-8-4007-0490-1/24/08}

\begin{document}

\title{Graph Bottlenecked Social Recommendation}
\author{Yonghui Yang}
\affiliation{
\institution{Hefei University of Technology}
\city{Hefei}
\country{China}
}
\email{yyh.hfut@gmail.com}

\author{Le Wu}
\authornotemark[1]
\affiliation{
\institution{Hefei University of Technology}
\city{Hefei}
\country{China}
}
\email{lewu.ustc@gmail.com}
\thanks{Le Wu is the Corresponding author}

\author{Zihan Wang}
\affiliation{
\institution{Hefei University of Technology}
\city{Hefei}
\country{China}
}
\email{zhwang.hfut@gmail.com}

\author{Zhuangzhuang He}
\affiliation{
\institution{Hefei University of Technology}
\city{Hefei}
\country{China}
}
\email{hyicheng223@gmail.com}

\author{Richang Hong}
\affiliation{
\institution{Hefei University of Technology}
\city{Hefei}
\country{China}
}
\email{hongrc.hfut@gmail.com}

\author{Meng Wang}
\affiliation{
\institution{Hefei University of Technology}
\city{Hefei}
\country{China}
}
\email{eric.mengwang@gmail.com}

\renewcommand{\shortauthors}{Yonghui Yang et al.}

\include{Alltex/0-abs}



\maketitle

\input{Alltex/1-intro}

\input{Alltex/3-prelimilaries}

\input{Alltex/4-model}

\input{Alltex/5-experiments}

\input{Alltex/2-related_works}

\input{Alltex/6-Conclusion}

\begin{acks}
This work was supported in part by grants from the National Key Research and Development Program of China( Grant No.2021ZD0111802), and the National Natural Science Foundation of China( Grant No. U23B2031, 721881011).
\end{acks}

\balance
\bibliographystyle{ACM-Reference-Format}
\bibliography{GBSR}
\end{document}

%% file: Alltex/0-abs.tex
\begin{abstract}
    With the emergence of social networks, social recommendation has become an essential technique for personalized services. Recently, 
    graph-based social recommendations have shown promising results by capturing the high-order social influence. Most empirical studies of graph-based social recommendations directly take the observed social networks into formulation, and produce user preferences based on social homogeneity. Despite the effectiveness, we argue that social networks in the real-world are inevitably noisy~(existing redundant social relations), which may obstruct precise user preference characterization. Nevertheless, identifying and removing redundant social relations is challenging due to a lack of labels. 
    In this paper, we focus on learning the denoised social structure to facilitate recommendation tasks from an information bottleneck perspective. Specifically, we propose a novel 
    \fullname~framework to tackle the social noise issue. \shortname~is a model-agnostic social denoising framework, that aims to maximize the mutual information between the denoised social graph and recommendation labels, meanwhile minimizing it between the denoised social graph and the original one. This enables \shortname~to learn the minimal yet sufficient social structure, effectively reducing redundant social relations and enhancing social recommendations. Technically, \shortname~consists of two elaborate components, preference-guided social graph refinement, and HSIC-based bottleneck learning. Extensive experimental results demonstrate the superiority of the proposed \shortname~, including high performances and good generality combined with various backbones. Our code is available at: https://github.com/yimutianyang/KDD24-GBSR.
\end{abstract}

%% file: Alltex/1-intro.tex
\section{Introduction}
Learning informative user and item representations is the key to building modern recommender systems. Classic collaborative filtering paradigm factorizes user-item interaction matrix to learn user and item representations, which is widely researched but usually limited by sparse interactions. With the proliferation of social media, social recommendation has become an important technique to provide personalized suggestions~\cite{tang2013social}. Both user-item interactions~\cite{shao2022faircf, shao2024average} and user-user social relations~\cite{DiffNet, wu2020diffnet++} are available on social platforms, prompting the development of various social recommendation methods designed to exploit these behavior patterns~\cite{ma2008sorec, konstas2009social}.

Following the social homophily~\cite{mcpherson2001birds} and social influence theory~\cite{marsden1993network}, many efforts are devoted to characterizing social relation effects on user preferences. Early works mainly focus on exploiting first-order social relations, i.e., social regularization that assumes socially connected users share similar preference~\cite{jamali2010matrix}, and social enhancement that incorporates user-trusted friends' feedback as auxiliary for the target user~\cite{guo2015trustsvd}. Recently, witnessed the power of graph neural networks~(GNNs) on machine learning~\cite{ICLR2017GCN, wu2020comprehensive, wu2020joint, cai2024mitigating, chen2023improving}, graph-based recommendations have attracted more and more attention~\cite{LightGCN, wu2020learning, yang2023generative, cai2024popularity}. Graph-based social recommendations~\cite{GraphRec, DiffNet, wu2022graph} achieve impressive progress in improving recommendation performances by formulating users' high-order interest propagation and social influence diffusion with GNNs. 

\begin{figure} [t]
    \centering
    \subfloat[Douban-Book]{
    \includegraphics[width=40mm]{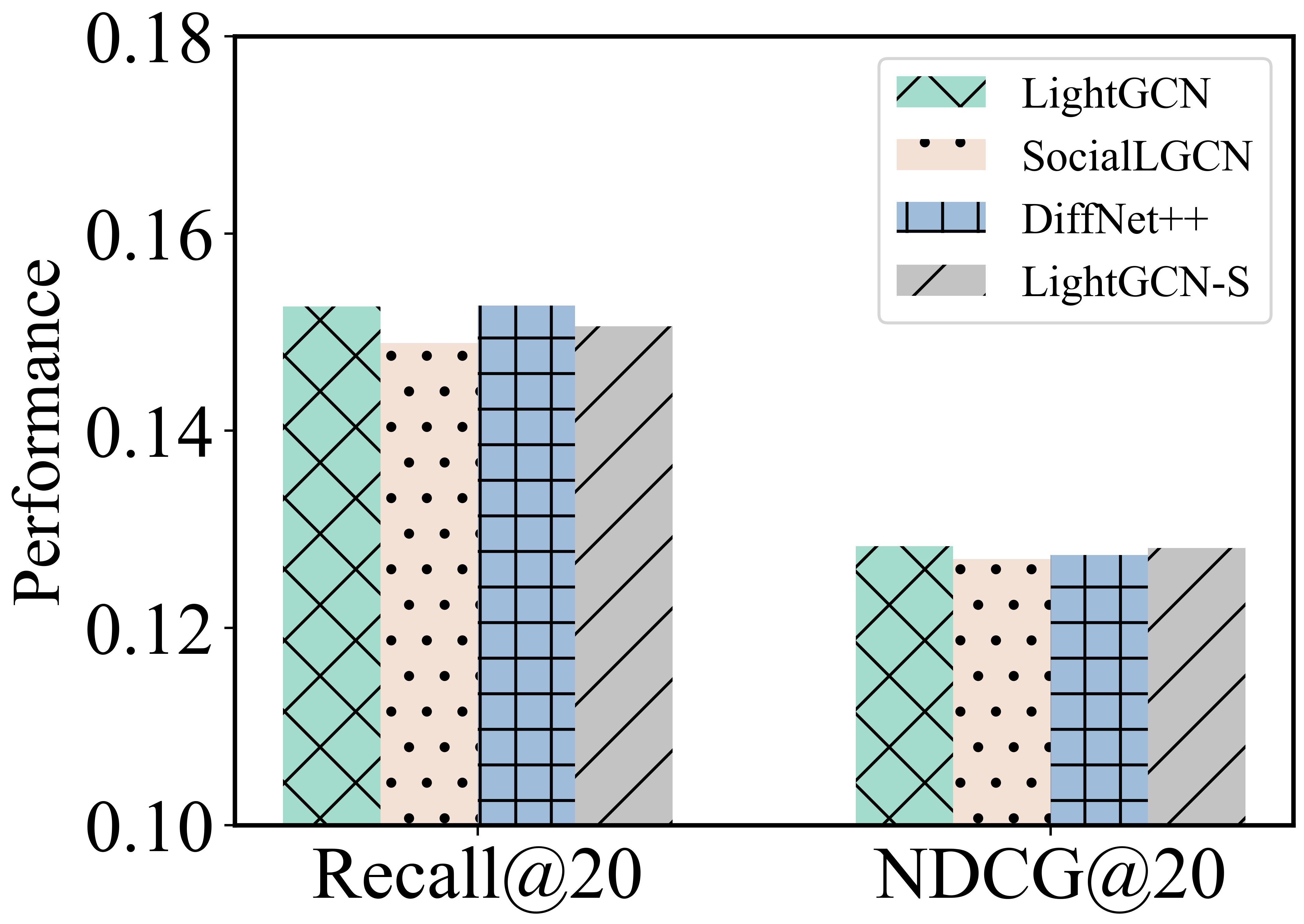}}
    \subfloat[Yelp]{\includegraphics[width=40mm]{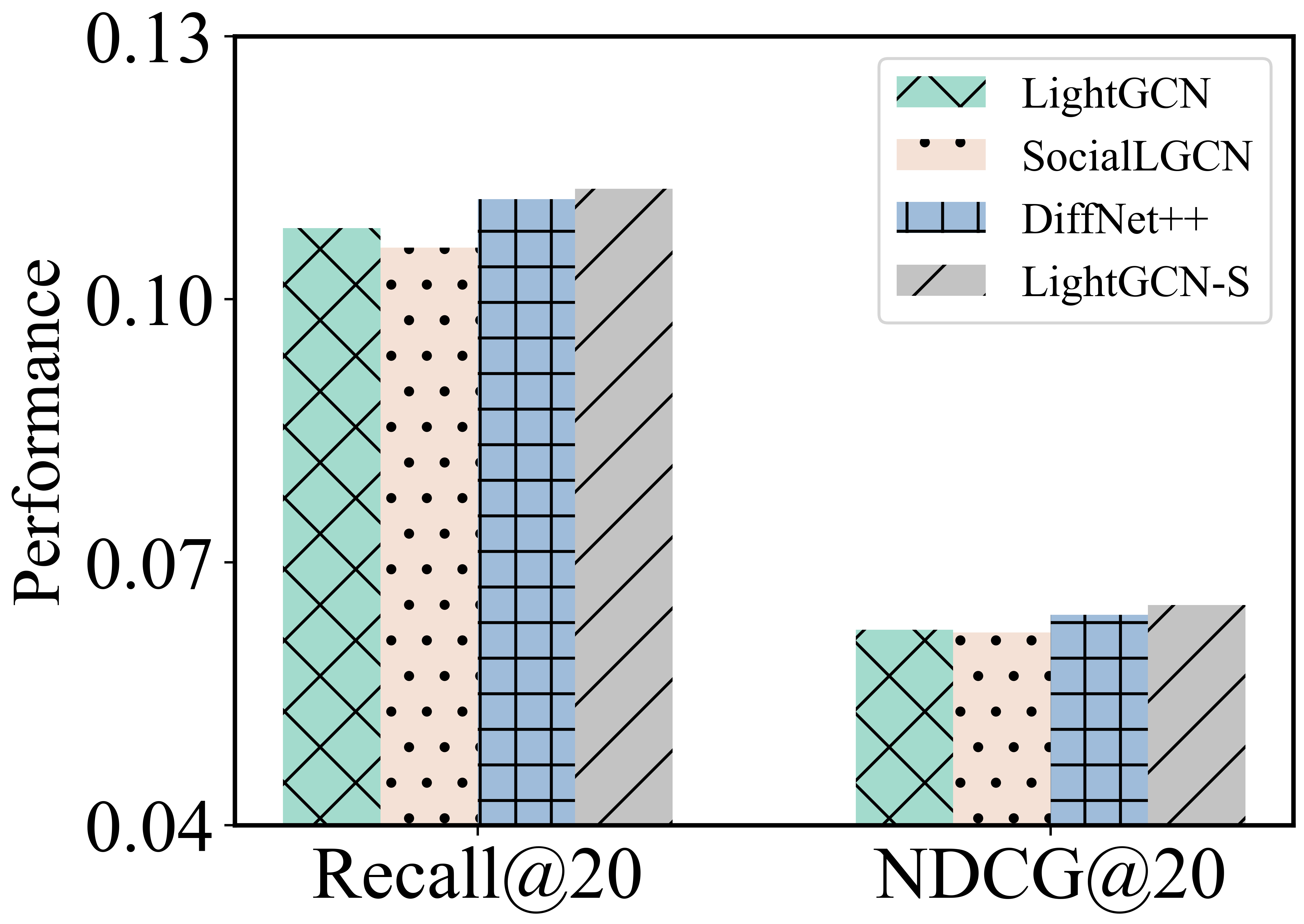}}
    \caption{Performance comparisons between LightGCN and SOTA graph-based social recommendation methods.}
    \label{fig: socialrec}
\end{figure}

Despite the effectiveness, current graph-based social recommendations rarely notice the social noise problem, i.e., social graphs are inevitably noisy with redundant social relations. Those redundant relations are caused by unreliable social relations and low preference-affinity social relations~\cite{sun2023denoising, WWW2023robust}. Consequently, directly using the observed social graph may hinder precise user preference characterization, leading to sub-optimal recommendation results. We conduct an empirical study to illustrate the social noise problem. As shown in Figure \ref{fig: socialrec}, we compare LightGCN with current SOTA graph-based recommendation methods, including SocialLGCN~\cite{liao2022sociallgn} and DiffNet++~\cite{wu2020diffnet++}. To avoid the effect of the message-passing mechanism of different methods, we additionally implement the extension of LightGCN, called LightGCN-S which additionally performs social neighbor aggregation for user representation learning. We can find that compared with LightGCN, graph-based social recommendations do not present significant strength on both metrics, even worse on the Douban-Book dataset. This indicates that social networks are usually noisy, it's necessary to filter redundant social relations to enhance the robustness of social recommendations. However, identifying and removing redundant social relations is non-trivial due to a lack of ground-truth labels. Besides, how can guarantee the recommendation accuracy while removing social relations?


In this paper, we focus on learning the denoised social graph structure to facilitate recommendation tasks from an information bottleneck perspective. Specifically, we propose a novel \fullname~framework to tackle the social noise problem. 
Let $\mathcal{G}^{S}=\{U, \mathbf{S}\}$ denote the user-user social graph and $\mathbf{R}$ denote the user-item interaction matrix, where $U$ is userset and $\mathbf{S}$ is social structure matrix. The optimal denoised social graph structure $\mathbf{S'}$ should satisfy:
\textit{the minimal from $\mathbf{S}$ yet efficient for infer $\mathbf{R}$.} To achieve this goal, we first introduce user preference signals to guide the social graph denoising process, then optimize the learning process via the Information Bottleneck~(IB) principle. Specifically, \shortname~maximize the mutual information between the denoised social graph structure $\mathbf{S'}$ and interaction matrix $\mathbf{R}$, meanwhile minimizing it between the denoised social graph structure $\mathbf{S'}$ and the original $\mathbf{S}$. Therefore, the learning objective is formulated as: $Max: I(\mathbf{R}; \mathbf{S'})-\beta I(\mathbf{S'}; \mathbf{S})$. 

Nevertheless, optimizing the objective of \shortname~for social recommendation is still challenging due to the following two challenges. For the maximization of $I(\mathbf{R}; \mathbf{S'})$, social graph and sparse interaction matrix are two non-Euclidean data, which are hard to compare directly.
For the minimization of $I(\mathbf{S'}; \mathbf{S})$, estimating the upper bound of MI is an intractable problem. Although some works~\cite{alemi2016VIB, cheng2020CLUB} leverage variational techniques to estimate the upper bound, they heavily rely on the prior assumption. To address the above two challenges, \shortname~is implemented as follows. First, regarding the hard-comparable issue of $I(\mathbf{R}; \mathbf{S'})$, we take all nodes into intermediary and derive the lower bound of $I(\mathbf{R}; \mathbf{S'})$ for maximization. Second, we introduce the Hilbert-Schmidt independence criterion~(HSIC)~\cite{ma2020HSIC} to replace the minimization of $I(\mathbf{S'}; \mathbf{S})$. HSIC~\cite{gretton2005measuring} is a statistic measure of 
variable dependency, minimizing HSIC approximate the minimization of mutual information.
Our contributions are summarized as follows:
\begin{itemize}
    \item In this paper, we revisit the social denoising recommendation from an information theory perspective, and propose a novel \fullname~framework to tackle the noise issue.
    \item Technically, we derive the lower bound of  $I(\mathbf{R}; \mathbf{S'})$ for maximization, and introduce the Hilbert-Schmidt independence criterion~(HSIC) to approximate the minimization of $I(\mathbf{S'}; \mathbf{S})$. 
    \item Empirical studies on three benchmarks clearly demonstrate the effectiveness and generality of the proposed \shortname~, i.e., \shortname~achives over $17.06\%$, $10\%$, and $11.27\%$ improvements of NDCG@20 compared with the strongest baseline.
\end{itemize}





%% file: Alltex/3-prelimilaries.tex
\section{Preliminaries}
\subsection{Problem Statement}
There are two kinds of entities in fundamental social recommendation scenarios: a userset $U$~($|U|=M$) and an itemset $V$~($|V|=N$). 
Users have two kinds of behaviors, user-user social relations and user-item interactions. We use matrix $\mathbf{S} \in \mathbb{R}^{M \times M}$ to describe user-user social structure, where each element $s_{ab}=1$ if user $b$ follows user $a$, otherwise $s_{ab}=0$. Similar, we use matrix {\small{$\mathbf{R} \in \mathbb{R}^{M\times N}$}} to describe user-item interactions, where each element $\mathbf{r}_{ai}=1$ if user $a$ interacted with item $i$, otherwise $\mathbf{r}_{ai}=0$. Given user $a$, item $i$, and social relation matrix $\mathbf{S}$ as input, graph-based social recommenders aim to infer the probability user $a$ will interact with item $i$: $\hat{r_{ai}}=\mathcal{G}_{\theta}(a,i,\mathbf{S})$, where $\mathcal{G}_{\theta}$ denotes GNN formulation. Thus, the optimization objective of graph-based social recommendation is defined as follows:
\begin{flalign}
    \theta^* = \mathop{\arg\min}\limits_{\theta}\mathbb{E}_{(a,i,r_{ai})\sim \mathcal{P}} \mathcal{L}_{r}(r_{ai}; \mathcal{G}_{\theta}(a,i,\mathbf{S})),
\end{flalign}

\noindent where $\mathcal{P}$ denote distribution of training data, and $\theta$ denote GNN parameters. However, user social networks are usually noisy with redundant relations~\cite{WWW2023robust}, directly using $\mathbf{S}$ to infer interaction probability may decrease the recommendation accuracy. In this work, we focus on learning robust social structure $\mathbf{S}'$ to facilitate recommendation performance:
\begin{flalign}
    \mathbf{S}'=\mathcal{F}_{\phi}(U, \mathbf{S}),
\end{flalign}

\noindent where $\mathcal{F}_{\phi}$ denotes social denoising function with the parameters $\phi$. Consequently, the final optimization of graph-noised social recommendation is described as follows:
\begin{flalign}
    \theta^*,\phi^* = \mathop{\arg\min}\limits_{\theta, \phi}\mathbb{E}_{(a,i,r_{ai})\sim \mathcal{P}} \mathcal{L}_{r}(r_{ai}; \mathcal{G}_{\theta}(a,i,\mathcal{F}_{\phi}(U, \mathbf{S}))).
\end{flalign}

\subsection{Information Bottleneck Principle}
Information Bottleneck~(IB) is a representation learning principle in machine learning, which seeks a trade-off between data fit and reducing irrelevant information~\cite{tishby2000information, tishby2015deep}. Given input data $X$, $Z$ is the hidden representation, and $Y$ is the downstream task label, which follows the Markov Chain $<X \rightarrow Z \rightarrow Y>$. IB principle describes that an optimal representation should maintain the minimal sufficient information for the downstream tasks~\cite{tishby2015deep, saxe2019information}: 
\begin{flalign}
    Z^{*} = \mathop{\arg\max}\limits_{Z} I(Y; Z)-\beta I(X; Z),
\end{flalign}

\noindent where $I(Y; Z)$ denotes the mutual information between the hidden representation $Z$ and label $Y$, $I(X; Z)$ denotes the mutual information between the hidden representation $Z$ and input data $X$
two variables, $\beta$ is the coefficient to balance these two parts. IB principle has been widely applied in machine learning tasks, such as model robustness~\cite{wu2020GIB, wang2021revisiting}, fairness~\cite{gronowski2023classification}, and explainability~\cite{bang2021explaining}. In this work, we introduce the IB principle to robust social denoising learning, which aims to seek the minimal yet sufficient social structure for recommendation tasks.

%% file: Alltex/4-model.tex
\section{The Proposed \shortname~Framework}
In this section, we introduce our proposed \fullname~framework for social denoising based recommendation. Essentially, \shortname~aims to learn the minimal yet efficient social structure to facilitate recommendation tasks, which is guaranteed by the information bottleneck principle. Next, we first give the overall optimization objective of \shortname~, then introduce how to implement each component of \shortname~in detail. Finally, we instantiate \shortname~with LightGCN-S backbone.

\subsection{Overview of \shortname}
\begin{figure}[t]
    \centering
    \includegraphics[width=82mm]{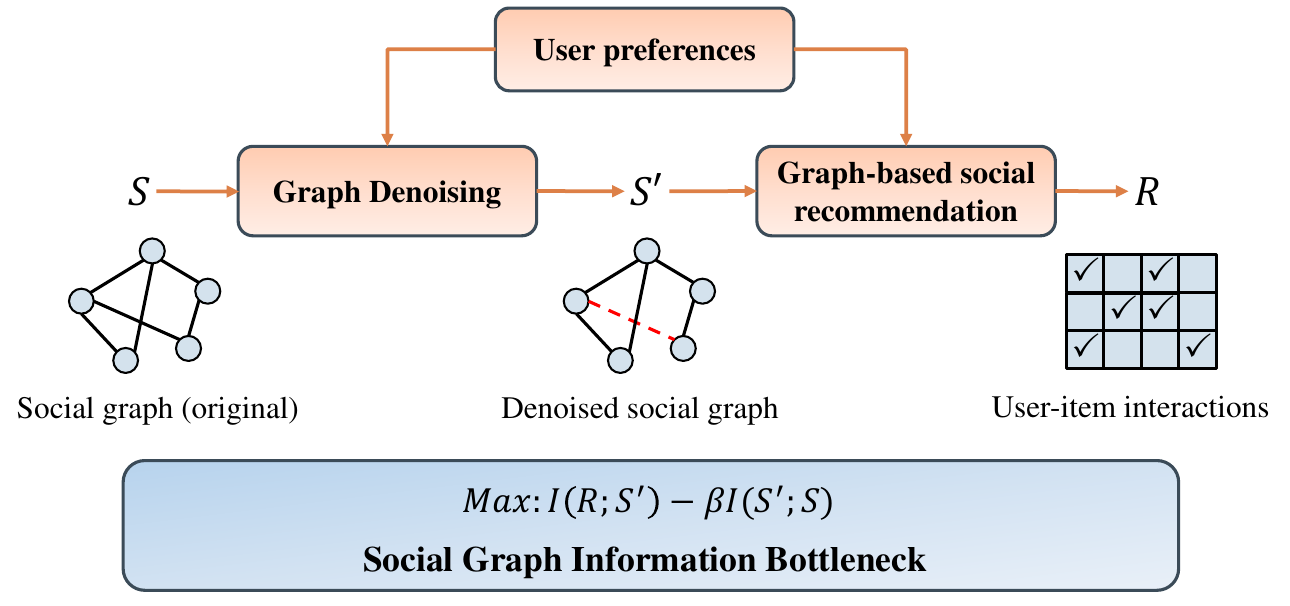}
    \vspace{-0.2cm}
    \caption{Overview of our proposed \shortname~framework.}
    \label{fig:intro}
\end{figure}

As shown in Figure \ref{fig:intro}, we present the overall objective of our proposed \shortname~framework for the social recommendation. Instead of directly using the original social structure $\mathbf{S}$, we aim to learn a denoised yet informative social structure $\mathbf{S}'$ to enhance recommendation. Due to the lack of available prior for social denoising, we introduce user preference signals to guide social graph denoising. To guarantee the trade-off between social denoising and recommendation tasks, we optimize \shortname~via graph information bottleneck principle. Thus, the goal of \shortname~is: $Max: I(\mathbf{R}; \mathbf{S'}) - \beta I(\mathbf{S'}; \mathbf{S})$. Due to the intractability of $I(\mathbf{R}; \mathbf{S'})$, we take all nodes into an intermediary for calculation. Thus, we obtain the final optimization objective of \shortname~:
\begin{flalign}
    Max: I(\mathbf{R}; U,V,\mathbf{S'}) -\beta I(\mathbf{S'}; \mathbf{S}), 
\end{flalign}

\noindent where the first term is encouraging that the denoised social graph preserves the essential information to facilitate recommendation tasks. The second term is the compression of the original social graph, aiming to filter redundant social relations.

\subsection{Preference-guided Social Denoising}
To achieve the above objective of \shortname~, we first need to refine the denoised social graph. The challenge is that although the social graph has noisy relations, there are no available labels to guide the denoising process. Based on social homogeneity social-connected individuals have more similar behavior similarity, we inject user preference signals into the social denoising process, i.e., users with similar preferences are more likely to have social relations.

Formally, we formulate the social denoising process as a graph edge dropout problem. Given the original social graph structure $\mathbf{S}$, the denoised one is defined as:
\begin{flalign} \label{eq: graph dropout}
    \mathbf{S'}= \mathcal{F}_{\phi}(U,\mathbf{S})=\{s_{ab} \odot \rho_{ab}\},
\end{flalign}

\noindent in which $\rho_{ab} \sim Bern(w_{ab})+\epsilon$ denotes that each edge $<u_a,u_b>$ will be dropped with the probability $1-w_{ab}+\epsilon$. Here, we add the parameter $\epsilon>0$ to represent the observation bias. 
Due to lacking prior information, we introduce task-relevant user preferences to refine social structure. 
Let $\mathbf{E}^{U} \in \mathbb{R}^{M \times d}$ denote user preference representations learned from the observed interactions, such as Matrix Factorization~\cite{PMF} and LightGCN~\cite{LightGCN}. For each observed social relation $<a,b>$, the link confidence is calculated as:
\begin{flalign} \label{eq: drop probability}
    w_{ab} = (g(\mathbf{e}_a, \mathbf{e}_b)),
\end{flalign}

\noindent where $\mathbf{e}_a$ and $\mathbf{e}_b$ denotes 
user $a$ and user $b$ preference representations, respectively. $g()$ is the fusion function, we employ MLPs to realize it. However, $\mathbf{S'}$ is not differentiable with the parameter $\rho$ of Bernoulli distribution, so we use the popular concrete relaxation method~\cite{jang2016categorical} to replace:
\begin{flalign} \label{eq: drop reparameter}
    Bern(w_{ab})=sigmoid({log(\delta/(1-\delta)+w_{ab})}/t),
\end{flalign}

\noindent where $\delta \sim U(0,1)$, and $t \in \mathbb{R^{+}}$ is the temperature parameter~(we set t=0.2 in our experiments). After re-parameterization, the discrete Bernoulli distribution is transferred to a differentiable function. 

\subsection{Maximization of $I(\mathbf{R}; U,V,\mathbf{S}')$}
Given the denoised social graph $\mathbf{S}'$, we first present how to maximize the mutual information $I(\mathbf{R}; U, V, \mathbf{S'})$, which ensures the denoised social graph satisfy recommendation tasks. Specifically, we derivate the lower bound of $I(\mathbf{R}; U,V, \mathbf{S'})$ as follows:
\begin{equation}
\begin{small}
    \begin{aligned}
         I(\mathbf{R}; U,V,\mathbf{S'}) & \overset{(a)}{=} H(\mathbf{R})-H(\mathbf{R}|U,V,\mathbf{S'})\\
    & \overset{(b)}{\geq} \sum_{a=0}^{M-1}\sum_{i=0}^{N-1}\sum_{r=0}^{1} p(r,a,i,\mathbf{S'})log(p(r|a,i,\mathbf{S'}) \\ 
    & \overset{(c)}{\geq} 
    \sum_{(a,i,j) \in \mathcal{D}}log(p(r_{ai}=1|a,i,\mathbf{S'}))+log(p(r_{ai}=0|a,j,\mathbf{S'})) \\
    & \overset{(d)}{=} \sum_{(a,i,j) \in \mathcal{D}}log(\sigma(\mathcal{G}(a,i,\mathbf{S'}))) - log(\sigma(\mathcal{G}(a,j,\mathbf{S'}))) \\
    & \overset{(e)}{\geq} \sum_{(a,i,j) \in \mathcal{D}}log(\sigma(\mathcal{G}(a,i,\mathbf{S'})-\mathcal{G}(a,j,\mathbf{S'}))),   
    \end{aligned}
    \end{small}
\end{equation}


\noindent where $\mathcal{G}(\cdot)$ is any graph-based social recommender as we mentioned in the preliminaries, $\sigma(\cdot)$is the sigmoid activation,
$\mathcal{D}=\{(a,i,j)|r_{ai}=1 \!\wedge\! r_{aj}=0\}$ is all training data. Next, we introduce each derivation step as follows: (a) is the definition of mutual information; (b) is the non-negative property of $H(\mathbf{R})$; (c) is that $p(r|a,i,\mathbf{S'})\leq 1$, and we split all samples into observed interactions and non-observed interactions; (d) $\sigma(\mathcal{G}(a,i,\mathbf{S'}))$ is the variational approximation of $p(r_{ai}=1|a,i,\mathbf{S'})$; (e) is due to $log(\sigma(x))-log(\sigma(y)) \geq log(\sigma(x-y))$. 

According to the above derivation, we can find that the popular BPR ranking loss~\cite{UAI2009BPR} is the lower bound of mutual information $I(\mathbf{R}; U,V, \mathbf{S'})$. Therefore, we employ BPR loss as the objective of mutual information maximization.


%


\subsection{Minimization of $I(\mathbf{S'}; \mathbf{S})$}
Next, we introduce how to minimize $I(\mathbf{S'}, \mathbf{S})$, which aims to reduce the redundant social relations in the original graph. Estimating the upper bound of mutual information is an intractable problem. Although some works~\cite{alemi2016VIB, cheng2020CLUB} leverage variational techniques to estimate the upper bound, but heavily rely on the prior assumption. Therefore, we introduce Hilbert-Schmidt Independence Criterion~(HSIC~\cite{gretton2005measuring}) as the approximation of the minimization of $I(\mathbf{R}; \mathbf{S'})$.

\textbf{HSIC brief.} HSIC serves as a statistical measure of dependency~\cite{gretton2005measuring}, which is formulated as the Hilbert-Schmidt norm, assessing the cross-covariance operator between distributions within the Reproducing Kernel Hilbert Space (RKHS). Mathematically, given two variables $X$ and $Y$, $\text{HSIC}(X, Y)$ is defined as follows:
\begin{small}
\begin{equation}
    \begin{aligned}
    HSIC(X, Y) &= ||C_{XY}||_{hs}^2 \\
    &= \mathbb{E}_{X,X',Y,Y'}[K_X(X,X')K_Y(Y,Y')] \\
    &+\mathbb{E}_{X,X'}[K_X(X,X')]\mathbb{E}_{Y,Y'}[K_Y(Y,Y')] \\
    &-2\mathbb{E}_{XY}[\mathbb{E}_{X'}[K_X(X,X')] \mathbb{E}_{Y'}[K_Y(Y,Y')]],
    \end{aligned}
    \label{eq: HSIC}
\end{equation}
\end{small}

\noindent where $K_X$ and $K_Y$ are two kernel functions for variables $X$ and $Y$, $X'$ and $Y'$ are two independent copies of $X$ and $Y$. Given the sampled instances ${(x_i, y_i)}_{i=1}^n$ from the batch training data, the $HSIC(X,Y)$ can be estimated as:
\begin{small}
\begin{equation}
    \hat{HSIC}(X,Y) = (n-1)^{-2}Tr(K_XHK_YH),
\end{equation}
\end{small}

\noindent where $K_X$ and $K_Y$ are used kernel matrices~\cite{gretton2005measuring}, with elements $K_{X_{ij}}=K_X(x_i,x_j)$ and $K_{Y_{ij}}=K_Y(y_i,y_j)$, $H=\mathbf{I}-\frac{1}{n}\mathbf{1}\mathbf{1}^T$ is the centering matrix, and $Tr(\cdot)$ denotes the trace of matrix. In practice, we adopt the widely used radial basis function~(RBF)~\cite{vert2004kernel} as the kernel function:
\begin{equation}
    \label{eq: kernel}
    K(x_i,x_j) = exp(-\frac{||x_i-x_j||^2}{2\sigma^2}),
\end{equation}
\noindent where $\sigma$ is the parameter that controls the sharpness of RBF. 

\textbf{HSIC-based bottleneck learning.} Given the original and denoised social graph structures $\mathbf{S}$ and $\mathbf{S'}$, we minimize $HSIC(\mathbf{S'}; \mathbf{S})$ to replace the minimization of $I(\mathbf{S'}; \mathbf{S})$. However, social graphs are non-Euclidean data, making it difficult to measure dependency. In practice, we adopt Monte Carlo sampling~\cite{shapiro2003monte} on all the node representations for calculation:
\begin{flalign} \label{eq: sampled hsic}
    Min: HSIC(\mathbf{S'}, \mathbf{S}) = \hat{HSIC}(\mathbf{E'}_{\mathcal{B}}, \mathbf{E}_{\mathcal{B}}),
\end{flalign}

\noindent where $\mathcal{B}$ denotes the batch sampling users, $\mathbf{E'}$ and $\mathbf{E}$ denote node representations, which are learned from recommenders $\mathcal{G}_{\theta, \phi}(U,V,\mathbf{S'})$ and $\mathcal{G}_{\theta}(U,V,\mathbf{S})$. Thus, we can reduce the redundant social relations via the HSIC-based bottleneck regularization:
\begin{flalign} \label{eq: hsic loss}
   \mathcal{L}_{ib} = \hat{HSIC}(\mathbf{E'}_{\mathcal{B}}, \mathbf{E}_{\mathcal{B}}).
\end{flalign}

\subsection{Instantiating the \shortname~Framework}
In this section, we instantiate our proposed \shortname~with specific graph-based social recommender $\mathcal{G}_{\theta}(U,V,\mathbf{S})$.  To avoid the effect of different message-passing mechanisms, we implement LightGCN-S as the backbone model~(we also realize \shortname~with other backbones, refer to the generality analysis). Firstly, we formulate the available data and denoised social structure as a graph $\mathcal{G}=\{U \cup V, \mathbf{A}\}$, where $U \cup V$ denotes the set of nodes, and $\mathbf{A}$ is the adjacent matrix defined as follows:
\begin{flalign}\label{eq:adj_matrix}
\mathbf{A}=\left[\begin{array}{cc}
\mathbf{S} & \mathbf{R}\\
\mathbf{R}^T & \mathbf{0}^{N\times N}
\end{array}\right].
\end{flalign}

\noindent Given the initialized node embeddings $\mathbf{E}^0 \in \mathbb{R}^{(M+N) \times d}$, LightGCN-S updates node embeddings through multiple graph convolutions:
\begin{flalign}\label{eq:aggregation}
\mathbf{E}^{l+1}
=\mathbf{D}^{-\frac{1}{2}}\mathbf{A}\mathbf{D}^{-\frac{1}{2}} \mathbf{E}^l, 
\end{flalign}

\noindent where $\mathbf{D}$ is the degree matrix of graph $\mathcal{G}$, $\mathbf{E}^{l+1}$ and $\mathbf{E}^{l}$ denote node embeddings in ${l+1}^{th}$ and ${l}^{th}$ graph convolution layer, respectively. When stacking $L$ graph convolution layers, the final node representations can be obtained with a readout operation:
\begin{flalign}\label{eq:readout}
\mathbf{E}&=Readout(\mathbf{E}^0, \mathbf{E}^1, ..., \mathbf{E}^L).
\end{flalign}

\noindent After obtaining the learned node representations through GCNs, LightGCN-S infers the propensity that user $a$ interacts with item $i$ by an inner product: $\hat{r}_{ai}=<e_a, e_i>$. All the above process are summarized as $\hat{r}_{ai}=\mathcal{G}_{\theta}(a,i,\mathbf{S})$. 

Next, we give the illustration of graph-denoised social recommendation. We first use the initialized node embeddings to obtain user preference representations $\mathbf{P}=\mathbf{E}^0[:M]$, then achieve the denoised social structure $\mathbf{S'}$ based on preference-guided social structure learning~(section 3.2). Given the learned denoised social structure $\mathbf{S'}$, we establish graph-denoised social recommender $\hat{r}_{ai}=\mathcal{G}_{\theta, \phi}(a,i,\mathbf{S'})$. Then, we select the pairwise ranking loss~\cite{UAI2009BPR} to optimize model parameters:
\begin{equation}
    \label{eq: loss_rec}
\mathcal{L}_{rec}=\sum_{a=0}^{M-1}\sum\limits_{(i,j)\in D_a }-log\sigma(\hat{r}_{ai}-\hat{r}_{aj}) + \lambda ||\mathbf{E}^0||^2,
\end{equation} 

\noindent where $\sigma(\cdot)$ is the sigmoid activation function, $\lambda$ is the regularization coefficient.
{\small$D_a=\{(i,j)|i\in R_a\!\wedge\!j\not\in R_a\}$} denotes the pairwise training data for user $a$. {\small$R_a$} represents user $a$ interacted items on the training data. Combined with the HSIC-based bottleneck regularizer, we obtain the final optimization objective:
\begin{flalign} \label{eq: all loss}
    \mathop{\arg\min}\limits_{\theta, \phi} \mathcal{L} = \mathcal{L}_{rec}+\beta \mathcal{L}_{ib},
\end{flalign}

\noindent The overall learning process of \shortname~is illustrated in Algorithm 1.

\begin{algorithm}[t]
\SetAlgoNlRelativeSize{0}
\caption{The algorithm of \shortname~}
\KwData{Userset $U$, Itemset $V$, User-item interactions $\mathbf{R}$, User-user social relations $\mathbf{S}$, and observation bias $\epsilon$}
\KwResult{Optimal graph-denoised social recommender $\mathcal{G}^{*}_{\theta, \phi}(\cdot)$}
Initialize recommender $\mathcal{G}_{\theta, \phi}$ with random weights\;

\While{not converged}{
    Sample a batch training data $\mathcal{D}$\;
    Compute social edge dropout probability via Eq.\eqref{eq: drop probability}-Eq.\eqref{eq: drop reparameter}\; 
    Refine the denoised social structure $\mathbf{S'}$ via Eq.\eqref{eq: graph dropout}\;
    Obtain node representations $\mathbf{E}'$ via $\mathcal{G}_{\theta, \phi}(U,V,\mathbf{S'})$\;
    Obtain node representations $\mathbf{E}$ via $\mathcal{G}_{\theta}(U,V,\mathbf{S})$\;
    Compute recommendation task loss $\mathcal{L}_{rec}$ via Eq.\eqref{eq: loss_rec}\;
    Compute HSIC bottleneck loss $\mathcal{L}_{ib}$ via Eq.\eqref{eq: hsic loss}\;
    Update model parameters according to Eq.\eqref{eq: all loss}\;
}
\label{alg: SGIB}
Return the optimal $\mathcal{G}^{*}_{\theta, \phi}(\cdot)$;\
\end{algorithm}

\subsection{Model Discussion}
In this section, we analyze the proposed \shortname~from model complexity and model generalization.

\subsubsection{Space Complexity}
As illustrated in Algorithm 1, the parameters of \shortname~are composed of two parts: graph-based social recommender parameters $\theta$ and social denoising parameters $\phi$. Among them, $\theta=\mathbf{E}^0$ are the general parameters equipped for backbone models~(such as LightGCN-S). $\phi$ are the parameters of MLPs, which are used to calculate the social edge confidence. Because $\phi$ are the shared parameters for all social edges, the additional parameters of \shortname~are ignorable compared with backbone models.

\subsubsection{Time Complexity}
Compared with the backbone model~(such as LightGCN-S), the additional time cost is social graph denoising and HSIC-bottleneck optimization. Social graph denoising is conducted on the observed social relations, which performs a sparse matrix. Besides, the time complexity of the HSIC-bottleneck regularizer lies in the number of the sampled nodes~(refer to Eq.\eqref{eq: sampled hsic}). In practice, we adopt a mini-batch training strategy to reduce the time cost of bottleneck learning, and the additional time cost of \shortname~is affordable.
Besides, as we remove redundant social relations, the denoised yet informative social graph makes \shortname~convergence much faster than the backbone model. Experiments also verify the efficiency of \shortname.


\subsubsection{Model Generalization}
The proposed \shortname~is designed for social denoising under graph-based social recommendation scenarios. It does not depend on specific graph-based social recommenders, such as DiffNet++~\cite{wu2020diffnet++} and SocialLGN~\cite{liao2022sociallgn}. Our proposed \shortname~is a flexible denoising framework to enhance social recommendations, we also conduct experiments on four backbones to demonstrate the generalization. Besides the backbone model, the idea of introducing the information bottleneck principle to graph denoising can also be generalized for different recommendation scenarios.

%% file: Alltex/5-experiments.tex
\section{Experiments}
In this section, we conduct extensive experiments on three real-world datasets to validate the effectiveness of our proposed \shortname~. We first introduce experimental settings, followed by recommendation performance comparisons. Finally, we give a detailed model investigation, including training efficiency, visualization of the denoised social graph, and parameter sensitivities.

\begin{table}[th]
    \centering
	\setlength{\belowcaptionskip}{5pt} %
	\caption{The statistics of three datasets.}\label{tab:statistics}
	\vspace{-0.2cm}
	\scalebox{1.0}{
    \begin{tabular}{c|c|c|c}
    \hline
    Dataset & Douban-Book & Yelp & Epinions \\ \hline
    Users & 13,024 & 19,593 & 18,202  \\ 
    Items & 22,347 & 21,266 & 47,449  \\ \hline
    Interactions & 792,062 & 450,884 & 298,173 \\ 
    Social Relations & 169,150 & 864,157 & 381,559 \\ \hline
    Interaction Density & 0.048\%  & 0.034\%  & 0.035\%  \\
    Relation Density & 0.268\% & 0.206\% & 0.115\% \\ \hline
    \end{tabular}}
    \vspace{-0.2cm}
\end{table}

\subsection{Experimental Settings}
\subsubsection{Datasets} We conduct empirical studies on three public datasets to verify the effectiveness of our proposed \shortname~, including Douban-Book, Yelp, and Epinions. All datasets contain user-user social links and user-item interactions. For the Douban-Book and Yelp datasets, we follow the released version in ~\cite{SEPT}. For the Epinions dataset, we follow the released version in ~\cite{HGSR}. Then, we sample 80\% interactions as training data, and the remaining 20\%  as test data. The detailed statistics of all datasets are summarized in Table \ref{tab:statistics}.

\begin{table*}[t]
\centering
\caption{Overall performance comparisons on three benchmarks. The best performance is highlighted in \textbf{bold} and the second is highlighted by \underline{underlines}. Impro. indicates the relative improvement of our proposed \shortname~compared to the best baseline.}
 \vspace{-2pt}
\label{tab: overall performance}
\scalebox{1.02}{
\begin{tabular}{|l|c|c|c|c|c|c|c|c|c|c|c|c|}
\hline 
 & \multicolumn{4}{c|}{Douban-Book}       
 & \multicolumn{4}{c|}{Yelp}
 & \multicolumn{4}{c|}{Epinions}\\ \cline{2-13}
\multirow{-2}{*}{Models} &R@10&N@10&R@20&N@20 &R@10&N@10&R@20&N@20 &R@10&N@10&R@20&N@20  \\ \hline
LightGCN  & 0.1039 & 0.1195 & 0.1526 & 0.1283 & 0.0698 & 0.0507 & 0.1081 & 0.0623 & 0.0432 & 0.0314 & 0.0675 & 0.0385 \\ \hline
GraphRec & 0.0971 & 0.1145 & 0.1453 & 0.1237 & 0.0672 & 0.0485 & 0.1077 & 0.0607 & 0.0436 & 0.0315 & 0.0681 & 0.0387 \\ \hline
DiffNet++ & 0.1010 & 0.1184 & 0.1489 & 0.1270 & 0.0707 & 0.0516 & 0.1114 & 0.0640 & 0.0468 & 0.0329 & 0.0727 & 0.0406 \\ \hline 
SocialLGN & 0.1034 & 0.1182 & 0.1527 & 0.1274 & 0.0681 & 0.0507 & 0.1059 & 0.0620 & 0.0416 & 0.0307 & 0.0634 & 0.0371 \\ \hline
LightGCN-S & 0.1021 & 0.1187 & 0.1506 & 0.1281 & 0.0714 & 0.0529 & 0.1126 & 0.0651 & \underline{0.0477} & \underline{0.0347} & 0.0716 & \underline{0.0417}  \\ \hline \hline
Rule-based & 0.1033 & 0.1192 & 0.1518 & 0.1289 & 0.0705 & 0.0526 & 0.1126 & 0.0652 & 0.0465 & 0.0340 & 0.0716 & 0.0414 \\ \hline 
ESRF & \underline{0.1042} & \underline{0.1199} & 0.1534 & \underline{0.1301} & 0.0718 & 0.0526 & 0.1123 & 0.0645 & 0.0462 & 0.0329 & \underline{0.0727} & 0.0406 \\ \hline
GDMSR  & 0.1026 & 0.1001 & \underline{0.1538} & 0.1245 & \underline{0.0739} & \underline{0.0535} & \underline{0.1148} & \underline{0.0658} & 0.0461 & 0.0326 & 0.0721 & 0.0414 \\ \hline


\textbf{GBSR}  & \textbf{0.1189} & \textbf{0.1451} & \textbf{0.1694} & \textbf{0.1523} & \textbf{0.0805} & \textbf{0.0592} & \textbf{0.1243} & \textbf{0.0724} & \textbf{0.0529} & \textbf{0.0385} & \textbf{0.0793} & \textbf{0.0464} \\ \hline
\textbf{Impro.} & \textbf{14.11\%} & \textbf{21.02\%} & \textbf{10.14\%} & \textbf{17.06\%} & \textbf{8.93\%} & \textbf{10.65\%} & \textbf{8.28\%} & \textbf{10.00\%} & \textbf{10.90\%} & \textbf{10.95\%} & \textbf{9.08\%} & \textbf{11.27\%}  \\ \hline
\end{tabular}}
\end{table*}

\subsubsection{Baselines and Evaluation Metrics.}
To evaluate the effectiveness of our proposed \shortname~, we select state-of-the-art baselines for comparisons. Specifically, these baselines can be divided into two groups: graph-based social recommendation methods~\cite{LightGCN,GraphRec,wu2020diffnet++} and social graph denoising methods~\cite{ICML2020robust, TKDE2020enhancing, WWW2023robust}, which are list as follows:
\begin{itemize}
    \item \textbf{LightGCN}~\cite{LightGCN}: is the SOTA graph-based collaborative filtering method, which simplifies GCNs by removing the redundant feature transformation and non-linear activation components for ID-based recommendation.
    \item \textbf{LightGCN-S}: We extend LightGCN to graph-based social recommendation, that each user's neighbors include their interacted items and linked social users. LightGCN-S is a basic and lightweight model, considering our proposed \shortname~ is a model-agnostic social graph denoising method, we select LightGCN-S as the backbone model.
    \item \textbf{GraphRec}~\cite{GraphRec}: is a classic graph-based social recommendation method, it incorporates user opinions and user two kinds of graphs for preference learning.
    \item \textbf{DiffNet++}~\cite{wu2020diffnet++}: is the SOTA graph-based social recommendation method, it recursively formulates user interest propagation and social influence diffusion process with a hierarchical attention mechanism.
    \item \textbf{SocialLGN}~\cite{liao2022sociallgn}: propagates user representations on both user-item interactions graph and user-user social graph with light graph convolutional layers, and fuses them for recommendation.
    \item \textbf{Rule-based}: We follow~\cite{WWW2023robust} and remove unreliable social relations based on the similarity of the user-interacted items.
    \item \textbf{ESRF}~\cite{TKDE2020enhancing}: generates alternative social neighbors and further performs neighbor denoising with adversarial training.
    \item \textbf{GDMSR}~\cite{WWW2023robust}:  designs the robust preference-guided social denoising to enhance graph-based social recommendation, it only remains the informative social relations according to preference confidences.
\end{itemize}

As we focus on implicit recommendation scenarios, we employ two widely used ranking metrics: Recall@N and NDCG@N~\cite{gunawardana2009survey, steck2013evaluation}. Specifically, Recall@N measures the percentage of the recalled positive samples for the Top-N ranking lists. Furthermore, NDCG@N assigns higher scores for those items in the top-ranked positions. In the evaluation stage, we adopt a full-ranking strategy that views all non-interacted items as candidates to avoid biased evaluation~\cite{krichene2020sampled, zhao2020revisiting}. For each model, we repeat experiments in 5 times and report the average values.

\subsubsection{Parameter Settings.} We implement our proposed \shortname~ and backbone with Tensorflow~\footnote{https://www.tensorflow.org}. For all baselines, we follow the original settings and carefully fine-tune parameters for fair comparisons. For latent embedding based methods, we initialize their embeddings with a Gaussian distribution with a mean value of 0 and a standard variance of 0.01, and fix the embedding size to 64. For model optimization, we use Adam optimizer with a learning rate of 0.001 and a batch size of 2048. We follow the mainstream ranking-based methods~\cite{UAI2009BPR}, and randomly select 1 non-interacted item as the negative sample for pairwise ranking optimization. We search the GCN layer in $[1,2,3,4]$, the regularization parameter $\lambda$ in $[0.0001, 0.001, 0.01]$. For the observation bias, we set $\epsilon=0.5$ for all datasets.
For information bottleneck constraint coefficient $\beta$, we use grid-search with different scales over three datasets, and report detailed analysis in experiments.

\subsection{Recommendation Performances}
\subsubsection{Overall Comparisons with Baselines.}
As shown in Table \ref{tab: overall performance}, we compare our proposed \shortname~with state-of-the-art methods on three benchmarks.  For a fair comparison, all denoising methods are conducted on the LightGCN-S backbone. Given the empirical studies, we have the following observations:
\begin{itemize}
    \item Compared with LightGCN, graph-based social recommendation methods present slight improvements under most of the datasets, i.e., DiffNet++ obtains a 2.24\% improvement on the NDCG@20 metric for Yelp dataset. However, this is not always the case, all social graph recommendations show a performance degradation on the Douban-Book dataset. While supported by social graphs, it is noteworthy that graph-based social recommendation methods do not consistently outperform LightGCN in terms of performance. These demonstrate that directly using social graphs may decrease recommendation performance, it's necessary to remove redundant social relations to enhance recommendation.
    
    \item Compared with directly using original social graphs, social denoising methods present better performances in most cases.
    This indicates that social noise is ubiquitous in real-world recommendation scenarios. All social denoising methods are implemented on LightGCN-S backbone, we find that GDMSR is the strongest baseline, which benefits from preference-guided social denoising and self-correcting curriculum learning. However, these social denoising methods don't present large-margin improvements compared with the backbone model. The reason is that simple rule or assumption based denoising methods lack of theoretical guarantee, it's hard to seek an effective trade-off between social denoising and recommendation accuracy.
    
    \item Our proposed \shortname~consistently outperforms all baselines under all experimental settings. Specifically, \shortname~ improves the strongest baseline $w.r.t$ NDCG@20 by 17.06\%, 10\% and 11.27\% on Douban-Book, Yelp, and Epinions datasets, respectively. Compared with the backbone model, \shortname~achieves impressive superiority over three benchmarks. These indicate that our proposed \shortname~ can significantly improve graph-based social recommendations, demonstrating the effectiveness of graph bottleneck learning to reduce redundant social relations. Compared with other social denoising methods, our \shortname~can better obtain the trade-off between removing social relations and recommendation tasks.
\end{itemize}

\begin{table*}[t]
\centering
\caption{Performance comparisons of \shortname~on different backbones.}
 \vspace{-5pt}
\label{tab: different backbones}
\scalebox{1.05}{
\begin{tabular}{l|c|c|c|c|c|c}
\hline 
 & \multicolumn{2}{c|}{Douban-Book}       
 & \multicolumn{2}{c|}{Yelp}
 & \multicolumn{2}{c}{Epinions}\\ \cline{2-7}
\multirow{-2}{*}{Models} & Recall@20&NDCG@20&Recall@20&NDCG@20&Recall@20&NDCG@20\\ \hline 
GraphRec & 0.1453 & 0.1237 & 0.1077 & 0.0607 & 0.0681 & 0.0387\\
+GBSR &0.1510(+3.92\%) & 0.1306(+5.58\%) & 0.1136(+5.48\%) & 0.0662(+9.06\%) & 0.0706(+3.67\%) & 0.0403(+4.13\%)\\ \hline
DiffNet++ & 0.1489 & 0.1270 & 0.1114 & 0.0640 & 0.0727 & 0.0406 \\
+GBSR & 0.1545(+3.76\%) & 0.1334(+5.04\%) & 0.1224(+9.87\%) & 0.0721(+12.66\%) & 0.0790+(+8.67\%) & 0.0451(+11.08\%)\\ \hline
SocialLGN & 0.1537 & 0.1274 & 0.1059 & 0.0620 & 0.0634 & 0.0371 \\
+GBSR &0.1591(+3.51\%) & 0.1353(+6.20\%) &0.1152(+8.78\%)&0.0675(+8.87\%)&0.0675(+6.47\%) &0.0399(+7.55\%) \\ \hline
LightGCN-S & 0.1506 & 0.1281  & 0.1126 & 0.0651 & 0.0716 & 0.0417 \\
+GBSR & 0.1694(+12.48\%) & 0.1523(+18.89\%) & 0.1243(+10.39\%) & 0.0724(+11.21\%) & 0.0793(+10.75\%) & 0.0464(+11.27\%) \\ \hline
\end{tabular}}
\end{table*}

\subsubsection{Ablation study}
We conduct ablation studies on three datasets to explore the effectiveness of each component of the proposed \shortname~framework. As shown in Table~\ref{tab: ablation study}, we compare \shortname~with corresponding variants on Top-20 recommendation performances. \textit{GBSR}-w/o HSIC denotes that remove the HSIC-based bottleneck regularization of \shortname~, we only keep preference-guided social denoising module. From Table~\ref{tab: ablation study}, we can find that GBSR-w/o HSIC performs worse in all cases, even worse than the backbone model. This indicates that simple social structure learning without HSIC-based bottleneck regularization is useless for recommendation tasks. Furthermore, under the constraint of the information bottleneck principle, the learned social structure is meaningful, which can effectively improve social recommendations on three datasets.

\subsubsection{Generality study of \shortname~}
As we mentioned in the model discussion, the proposed \shortname~is a model-agnostic social denoising framework. 
To better illustrate the generality of \shortname~, we conduct experiments of \shortname~on several graph-based social recommendation backbones. As shown in Table \ref{tab: different backbones}, we implement \shortname~under four backbones, including GraphRec~\cite{GraphRec}, DiffNet++\cite{wu2020diffnet++}, SocialLGN~\cite{liao2022sociallgn}, and LightGCN-S, and report their performances of Top-20 recommendation task.
From Table \ref{tab: different backbones}, we observe that \shortname~consistently outperforms each backbone by a large margin. For example, on the Yelp dataset, \shortname~achieves 9.06\%, 12.66\%, 8.87\%, and 11.21\% improvements of NDCG@20 compared with GraphRec, DiffNet++, SocialLGN, and LightGCN-S, respectively. Similarly, \shortname~also obtains 5.48\%, 9.87\%, 8.78\%, and 10.39\% improvements on the Recall@20 metric.
Extensive experimental results show that our proposed \shortname~has a good generalization ability, which can easily coupled with current graph-based social recommendation methods and further enhancement.

\subsection{Investigation of \shortname~}
In this section, we further analyze \shortname~from the following aspects: training efficiency, visualization of the denoised social graphs, and hyper-parameter sensitivity analysis.

\begin{table}[t]
\centering
\caption{Ablation study on three datasets.}
 \vspace{-5pt}
\label{tab: ablation study}
\scalebox{0.90}{
\begin{tabular}{|l|c|c|c|c|c|c|}
\hline 
 & \multicolumn{2}{c|}{Douban-Book}       
 & \multicolumn{2}{c|}{Yelp}
 & \multicolumn{2}{c|}{Epinions}\\ \cline{2-7}
\multirow{-2}{*}{Models} & R@20&N@20&R@20&N@20&R@20&N@20\\ \hline 
LightGCN-S & 0.1506 & 0.1281 & 0.1126 & 0.0651 & 0.0716 & 0.0417\\ \hline
\textit{GBSR}-w/o HSIC & 0.1482 & 0.1259 & 0.1119 & 0.0644 & 0.0688 & 0.0388\\ \hline
\textit{GBSR} & 0.1694 & 0.1523 & 0.1243 & 0.0724 & 0.0793 & 0.0464\\ \hline
\end{tabular}}
\end{table}

\begin{figure}[t]
    \centering
    \includegraphics[width=42mm]{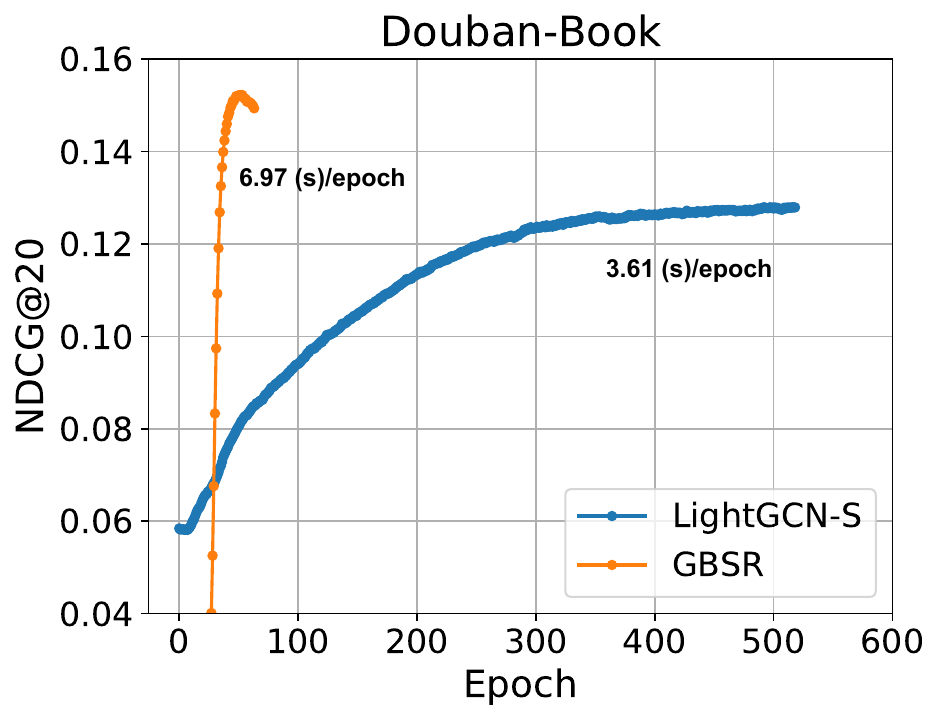}
    \includegraphics[width=42mm]{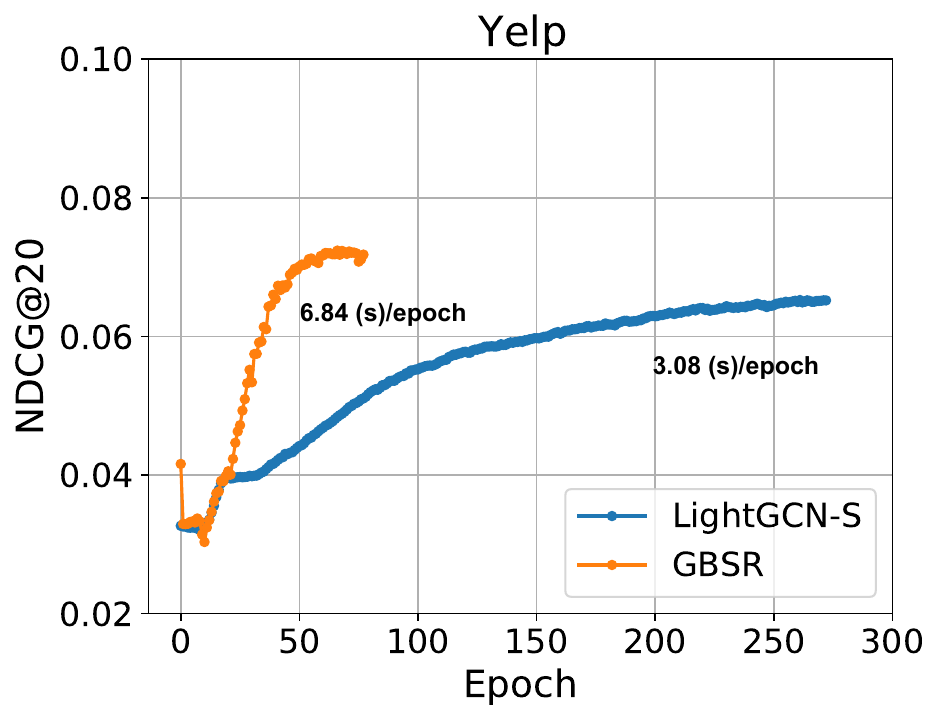}
    \caption{Convergence curves of \shortname~and LightGCN-S on Douban-Book and Yelp datasets.}
    \label{fig:training efficiency}
\end{figure}

\subsubsection{Training efficiency of GBSR}
To analyze the training efficiency of \shortname~, we compare the convergence speed of \shortname~and corresponding backbone~(LightGCN-S).
As shown in Figure \ref{fig:training efficiency}, we compare the convergence process of both models. As the space limit, we only present the convergence process on Douban-Book and Yelp datasets. We set gcn layer to 3 and keep all experimental settings the same. According to these figures, we can observe that \shortname~converges much faster than the backbone model. Particularly, \shortname~reaches the best performances at the $82^{th}$, the $67^{th}$ epoch on Douban-Book and Yelp datasets. In contrast, LightGCN-S obtains the best results on $509^{th}$, and $261^{th}$ epoch, respectively. Empirical evidence shows that \shortname~convergence 2-3 times faster than LightGCN-S.

\subsubsection{Visualization and statistics of the denoised social graphs}
\begin{figure}[t]
    \centering
    \subfloat[Ego-network~(from Douban-Book)]{\includegraphics[width=38mm]{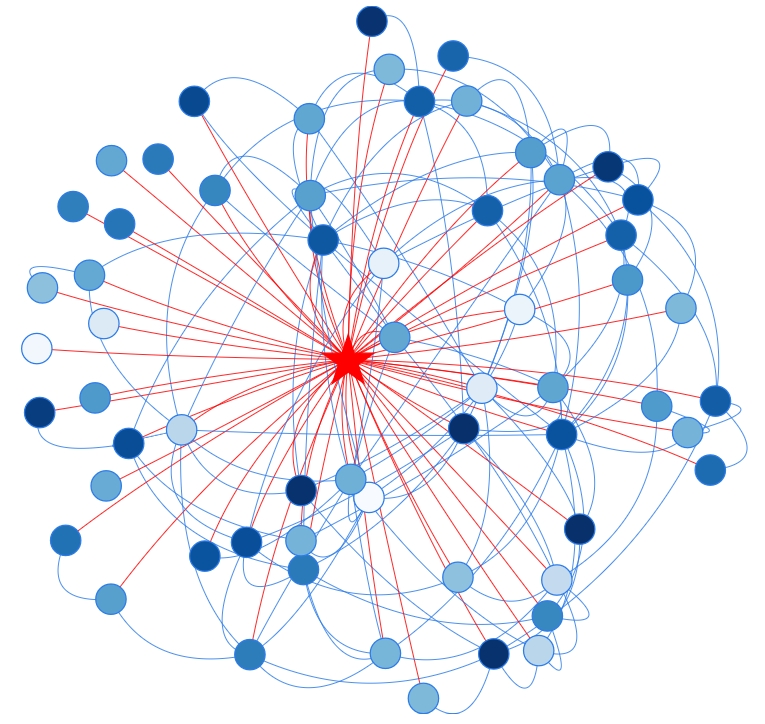}}
    \subfloat[Errorbar of social relation confidence]{
    \includegraphics[width=40mm]{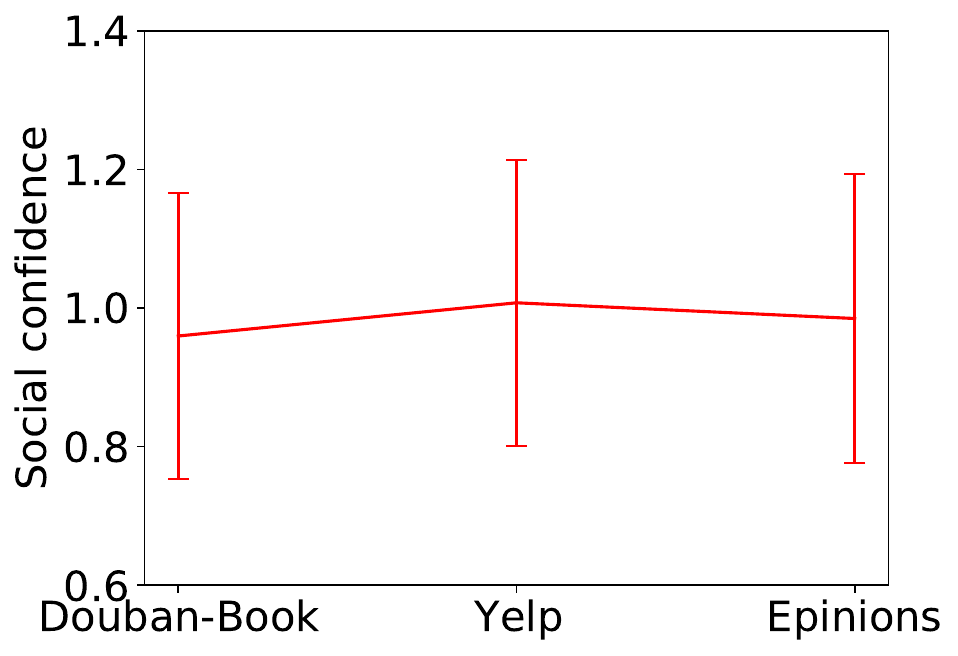}
    }
    \caption{Visualization and statistics of the denoised social graphs.}
    \label{fig: visualization}
\end{figure}

Here we first present the visualization of the denoised social graph.
As shown in Figure \ref{fig: visualization}(a), we present the sampled ego-network from Douban-Book datasets. The red node denotes the center user of this ego-network, and the blue nodes denote social neighbors. The depth of the node color denotes the probability of edge dropping,  where the darker the color, the lower the dropping probability. We can observe that user social neighbors perform different confidences of social relations.  Besides, we analyze the statistics of the denoised social graphs. As shown in Figure \ref{fig: visualization}(b), we plot the mean and variance values of social relation confidence on three datasets. We can observe that Douban-Book presents the lowest mean value of social confidence, which means that it has the most social noise over the three datasets. This also explains the results of Figure \ref{fig: socialrec} that graph-based social recommendations show a performance decrease compared with LightGCN on the Douban-Book dataset. These results demonstrate that the proposed \shortname~can effectively refine the observed social graph via information bottleneck, which provides informative social structures to enhance social recommendations.

\subsubsection{Parameter Sensitivity Analysis.} 
\begin{figure*}[th]
    \centering
    \subfloat[Douban-Book dataset]{\includegraphics[width=56mm]{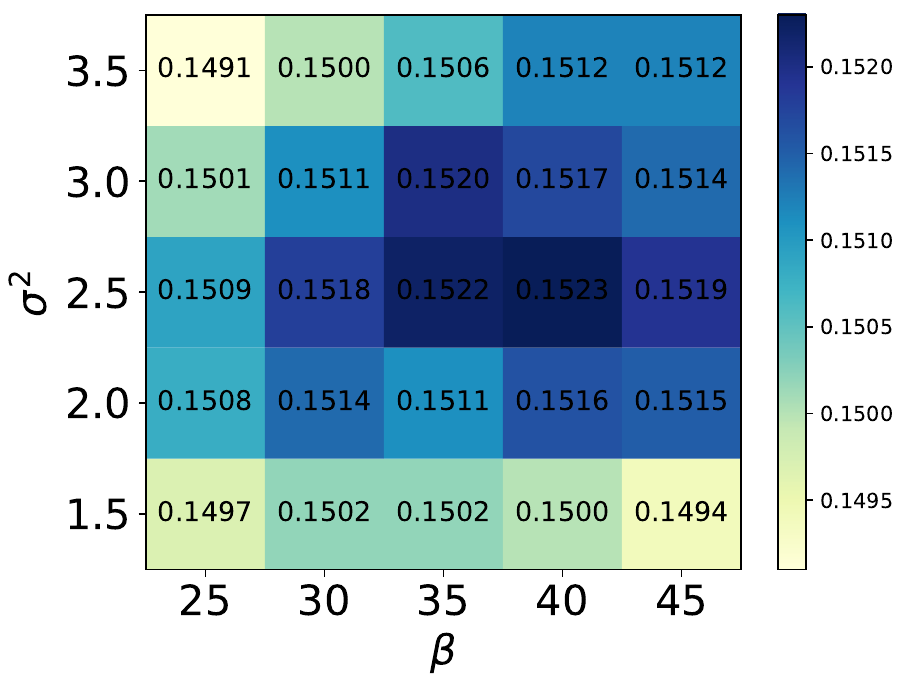}}
    \subfloat[Yelp dataset]{\includegraphics[width=56mm]{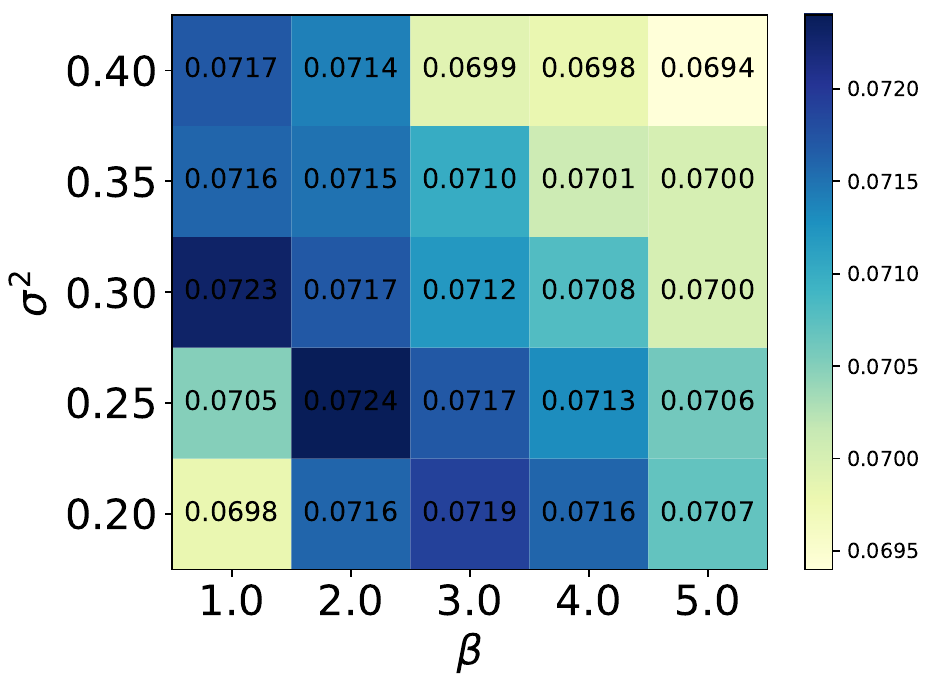}}
    \subfloat[Epinions dataset]{\includegraphics[width=56mm]{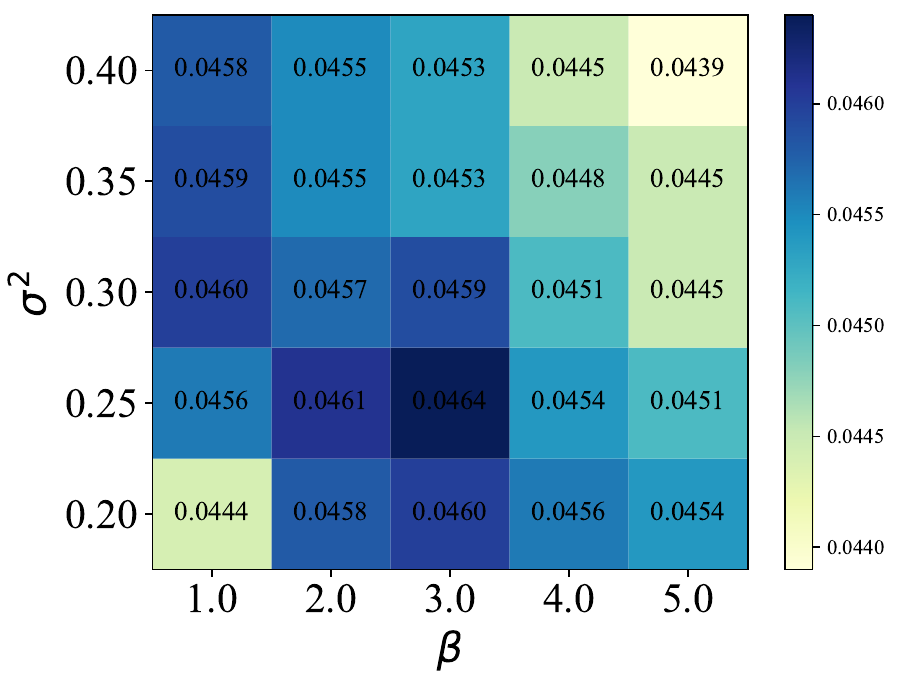}}
    \caption{Performance comparisons under different parameters $(\beta, \sigma^2)$.}
    \label{fig: parameter sensitivity}
\end{figure*}

In this part, we analyze the impact of different hyper-parameters of \shortname~. There are two key parameters, bottleneck loss coefficient $\beta$ and RBF sharpness parameter $\sigma^2$. As both parameters determine the scale of bottleneck loss, we combine them to analyze the influence of recommendation results. As shown in Figure \ref{fig: parameter sensitivity}, we conduct careful grid-search of $(\beta, \sigma^2)$ on three datasets. We can observe that \shortname~reaches the best performance when $\beta=40, \sigma^2=2.5$ on Douban-Book, $\beta=2.0, \sigma^2=0.25$ on Yelp, and $\beta=3.0, \sigma^2=0.25$, respectively.

%% file: Alltex/2-related_works.tex
\section{Related Works}

\subsection{Graph-based Social Recommendation}
With the emergence of social media, social recommendation has been an important technique and has attracted more and more research attention~\cite{tang2013social, konstas2009social, ma2008sorec, jamali2010matrix, ma2011recommender}. Following the social homophily~\cite{mcpherson2001birds} and social influence theory~\cite{marsden1993network}, social recommendations are devoted to characterizing social relation effects on user preferences. Early efforts exploit social relations in a shallow form, such as co-factorization methods~\cite{ma2008sorec, konstas2009social} and regularization-based methods~\cite{jamali2010matrix, ma2011recommender, jiang2014scalable}. For example, SoRec~\cite{ma2008sorec} jointly co-factorize the interaction and social matrices and then project interaction and social contexts into the same semantic space. \cite{ma2011recommender} designs a social regularization term that assumes two socially connected users should be closer in preference space. 
Recently, with the great success of graph neural networks~\cite{ICLR2017GCN, ICLR2018GAT}, graph-based social recommendations have been widely researched and achieved impressive process~\cite{GraphRec, DiffNet, wu2020diffnet++, liao2022sociallgn, yu2021self, HGSR}. By formulating user-user social relations as a graph, graph-based social recommendations inject high-order social influences into user preference learning, vibrant the representation ability. For example, DiffNet models the high-order social influence diffusion process to enhance user representation~\cite{DiffNet}, and DiffNet++ further improves it by combining both social influence diffusion and user-item interest propagation with a hierarchical attention mechanism~\cite{wu2020diffnet++}. Inspired by the architecture of LightGCN, \cite{liao2022sociallgn} proposes SocialLGN to model user interaction and social behaviors. Instead of learning social graphs on Euclidean space, some works attempt to introduce hyperbolic learning for graph-based social recommendations~\cite{TOIS2021hypersorec, HGSR}. 
Despite the effectiveness of modeling high-order social influence to improve recommendation, these works are built on the clean social relation assumption. However, social graphs are inevitably noisy with redundant relations, and these graph-based social recommendation methods are usually far from satisfactory. Instead of directly using the original social graph, in this work, we propose a graph noising framework to improve social recommendation.

\subsection{Recommendation Denoising}
Recommendation denoising works mainly focus on implicit feedback, which aims to refine implicit feedback to build robust recommender systems~\cite{wang2021denoising, yang2021enhanced, wang2023efficient, gao2022self, he2024double}. Most efforts are devoted to removing noise feedback, which is easily vulnerable to users' unconscious behaviors and various biases. For example, \cite{wang2021denoising} proposes to drop noisy feedback based on the observation that noisy feedback has higher training loss, \cite{wang2023efficient} devises a bi-level optimization method to implement recommendation denoising. Besides, graph augmentation methods are proposed to realize recommendation denoising~\cite{yang2021enhanced, fan2023graph}. 
Different from the above feedback-based denoising works, we focus on social denoising for recommendations. Social graphs are inevitably noisy with redundant relations, including unreliable relations and low preference-affinity relations~\cite{WWW2023robust, sun2023denoising}.
Early works employ statistics to identify unstable social relations~\cite{ma2011recommender, pan2020correlative}, or model different user influences with attention mechanism~\cite{sun2018attentive, wu2020diffnet++}. Besides, fine-grained social leveraging~\cite{fu2021dual} and adversarial learning based methods have been proposed~\cite{yu2019generating, TKDE2020enhancing}. Recently, GDMSR~\cite{WWW2023robust} proposes a distilled social graph based on progressive preference-guided social denoising. 
Nevertheless, the above methods still face the challenge of lacking ground-truth. Whether rule-based or assumption-based social denoising is hard to guarantee the trade-off between social denoising and social recommendation. Distinguished by these denoising methods, we address the social denoising recommendation from a novel information bottleneck perspective, which seeks the denoised yet informative social structure to enhance recommendations.

\subsection{Information Bottleneck and Applications}
Information Bottleneck~(IB) is an effective representation learning principle in machine learning tasks, that the optimal representation should satisfy the minimal yet efficient manner~\cite{tishby2000information, tishby2015deep}. In the era of deep learning, calculating high-dimensional variables' mutual information~(MI) is the key challenge for IB. The general solution is estimating the upper/lower bounds instead of directly calculating mutual information~\cite{alemi2016VIB, cheng2020CLUB}. Specifically, VIB~\cite{alemi2016VIB} leverages the variational technique to estimate the bounds of mutual information. Besides, MINE~\cite{belghazi2018mine}, InfoNCE~\cite{oord2018InfoNCE} are proposed to estimate the lower bound of MI. In contrast, a few attempts propose to estimate the upper bound of MI~\cite{cheng2020CLUB, hledik2019tight}. Besides optimizing the bounds of MI, HSIC-based methods~\cite{ma2020HSIC, wang2021revisiting} are proposed to implement IB learning, which employs the Hilbert-Schmidt Independence Criterion~(HSIC) to replace mutual information for optimization. HSIC measures the independence of two variables, which can approximate the mutual information objective~\cite{gretton2005measuring}.
IB principle has been successfully applied to many applications, such as image classification~\cite{wang2023learning}, text understanding~\cite{patrick2020support}, and graph learning~\cite{wu2020GIB}. 
In this work, we introduce the HSIC-based bottleneck to the graph-denoised social recommendation, aiming to filtering redundant social relations for robust recommendation.

%% file: Alltex/6-Conclusion.tex
\section{Conclusion}
In this paper, we investigate graph-denoised social recommendations and propose a novel \fullname~framework. Specifically, \shortname~aims to learn the denoised yet informative social structure for recommendation tasks. To achieve this goal, we first design preference-guided social denoising, then optimize the denoising process via the information bottleneck principle. Particularly, we derive the lower bound of mutual information maximization and introduce HSIC regularization to replace mutual information minimization. Extensive experiments conducted on three benchmarks demonstrate the effectiveness of our proposed \shortname~framework, i.e., over 10\% improvements on Top-20 Recommendation. Moreover, \shortname~is a model-agnostic framework, which can be flexibly coupled with various graph-based social recommenders. In the future, we will explore more potential of leveraging the IB principle to recommendation tasks, i.e., self-supervised recommendation, fairness-aware recommendation, and LLM-enhanced recommendation.

%% file: main.bbl

\begin{thebibliography}{69}


\ifx \showCODEN    \undefined \def \showCODEN     #1{\unskip}     \fi
\ifx \showDOI      \undefined \def \showDOI       #1{#1}\fi
\ifx \showISBNx    \undefined \def \showISBNx     #1{\unskip}     \fi
\ifx \showISBNxiii \undefined \def \showISBNxiii  #1{\unskip}     \fi
\ifx \showISSN     \undefined \def \showISSN      #1{\unskip}     \fi
\ifx \showLCCN     \undefined \def \showLCCN      #1{\unskip}     \fi
\ifx \shownote     \undefined \def \shownote      #1{#1}          \fi
\ifx \showarticletitle \undefined \def \showarticletitle #1{#1}   \fi
\ifx \showURL      \undefined \def \showURL       {\relax}        \fi
\providecommand\bibfield[2]{#2}
\providecommand\bibinfo[2]{#2}
\providecommand\natexlab[1]{#1}
\providecommand\showeprint[2][]{arXiv:#2}

\bibitem[Alemi et~al\mbox{.}(2017)]%
        {alemi2016VIB}
\bibfield{author}{\bibinfo{person}{Alexander~A Alemi}, \bibinfo{person}{Ian Fischer}, \bibinfo{person}{Joshua~V Dillon}, {and} \bibinfo{person}{Kevin Murphy}.} \bibinfo{year}{2017}\natexlab{}.
\newblock \showarticletitle{Deep variational information bottleneck}. In \bibinfo{booktitle}{\emph{ICLR}}.
\newblock


\bibitem[Bang et~al\mbox{.}(2021)]%
        {bang2021explaining}
\bibfield{author}{\bibinfo{person}{Seojin Bang}, \bibinfo{person}{Pengtao Xie}, \bibinfo{person}{Heewook Lee}, \bibinfo{person}{Wei Wu}, {and} \bibinfo{person}{Eric Xing}.} \bibinfo{year}{2021}\natexlab{}.
\newblock \showarticletitle{Explaining a black-box by using a deep variational information bottleneck approach}. In \bibinfo{booktitle}{\emph{AAAI}}, Vol.~\bibinfo{volume}{35}. \bibinfo{pages}{11396--11404}.
\newblock


\bibitem[Belghazi et~al\mbox{.}(2018)]%
        {belghazi2018mine}
\bibfield{author}{\bibinfo{person}{Mohamed~Ishmael Belghazi}, \bibinfo{person}{Aristide Baratin}, \bibinfo{person}{Sai Rajeshwar}, \bibinfo{person}{Sherjil Ozair}, \bibinfo{person}{Yoshua Bengio}, \bibinfo{person}{Aaron Courville}, {and} \bibinfo{person}{Devon Hjelm}.} \bibinfo{year}{2018}\natexlab{}.
\newblock \showarticletitle{Mutual information neural estimation}. In \bibinfo{booktitle}{\emph{ICML}}. PMLR, \bibinfo{pages}{531--540}.
\newblock


\bibitem[Cai et~al\mbox{.}(2024a)]%
        {cai2024popularity}
\bibfield{author}{\bibinfo{person}{Miaomiao Cai}, \bibinfo{person}{Lei Chen}, \bibinfo{person}{Yifan Wang}, \bibinfo{person}{Haoyue Bai}, \bibinfo{person}{Peijie Sun}, \bibinfo{person}{Le Wu}, \bibinfo{person}{Min Zhang}, {and} \bibinfo{person}{Meng Wang}.} \bibinfo{year}{2024}\natexlab{a}.
\newblock \showarticletitle{Popularity-Aware Alignment and Contrast for Mitigating Popularity Bias}.
\newblock \bibinfo{journal}{\emph{arXiv preprint arXiv:2405.20718}} (\bibinfo{year}{2024}).
\newblock


\bibitem[Cai et~al\mbox{.}(2024b)]%
        {cai2024mitigating}
\bibfield{author}{\bibinfo{person}{Miaomiao Cai}, \bibinfo{person}{Min Hou}, \bibinfo{person}{Lei Chen}, \bibinfo{person}{Le Wu}, \bibinfo{person}{Haoyue Bai}, \bibinfo{person}{Yong Li}, {and} \bibinfo{person}{Meng Wang}.} \bibinfo{year}{2024}\natexlab{b}.
\newblock \showarticletitle{Mitigating Recommendation Biases via Group-Alignment and Global-Uniformity in Representation Learning}.
\newblock \bibinfo{journal}{\emph{ACM Transactions on Intelligent Systems and Technology}} (\bibinfo{year}{2024}).
\newblock


\bibitem[Chen et~al\mbox{.}(2023)]%
        {chen2023improving}
\bibfield{author}{\bibinfo{person}{Lei Chen}, \bibinfo{person}{Le Wu}, \bibinfo{person}{Kun Zhang}, \bibinfo{person}{Richang Hong}, \bibinfo{person}{Defu Lian}, \bibinfo{person}{Zhiqiang Zhang}, \bibinfo{person}{Jun Zhou}, {and} \bibinfo{person}{Meng Wang}.} \bibinfo{year}{2023}\natexlab{}.
\newblock \showarticletitle{Improving recommendation fairness via data augmentation}. In \bibinfo{booktitle}{\emph{WWW}}. \bibinfo{pages}{1012--1020}.
\newblock


\bibitem[Cheng et~al\mbox{.}(2020)]%
        {cheng2020CLUB}
\bibfield{author}{\bibinfo{person}{Pengyu Cheng}, \bibinfo{person}{Weituo Hao}, \bibinfo{person}{Shuyang Dai}, \bibinfo{person}{Jiachang Liu}, \bibinfo{person}{Zhe Gan}, {and} \bibinfo{person}{Lawrence Carin}.} \bibinfo{year}{2020}\natexlab{}.
\newblock \showarticletitle{Club: A contrastive log-ratio upper bound of mutual information}. In \bibinfo{booktitle}{\emph{ICML}}. PMLR, \bibinfo{pages}{1779--1788}.
\newblock


\bibitem[Fan et~al\mbox{.}(2019)]%
        {GraphRec}
\bibfield{author}{\bibinfo{person}{Wenqi Fan}, \bibinfo{person}{Yao Ma}, \bibinfo{person}{Qing Li}, \bibinfo{person}{Yuan He}, \bibinfo{person}{Eric Zhao}, \bibinfo{person}{Jiliang Tang}, {and} \bibinfo{person}{Dawei Yin}.} \bibinfo{year}{2019}\natexlab{}.
\newblock \showarticletitle{Graph neural networks for social recommendation}. In \bibinfo{booktitle}{\emph{WWW}}. \bibinfo{pages}{417--426}.
\newblock


\bibitem[Fan et~al\mbox{.}(2023)]%
        {fan2023graph}
\bibfield{author}{\bibinfo{person}{Ziwei Fan}, \bibinfo{person}{Ke Xu}, \bibinfo{person}{Zhang Dong}, \bibinfo{person}{Hao Peng}, \bibinfo{person}{Jiawei Zhang}, {and} \bibinfo{person}{Philip~S Yu}.} \bibinfo{year}{2023}\natexlab{}.
\newblock \showarticletitle{Graph collaborative signals denoising and augmentation for recommendation}. In \bibinfo{booktitle}{\emph{SIGIR}}. \bibinfo{pages}{2037--2041}.
\newblock


\bibitem[Fu et~al\mbox{.}(2021)]%
        {fu2021dual}
\bibfield{author}{\bibinfo{person}{Bairan Fu}, \bibinfo{person}{Wenming Zhang}, \bibinfo{person}{Guangneng Hu}, \bibinfo{person}{Xinyu Dai}, \bibinfo{person}{Shujian Huang}, {and} \bibinfo{person}{Jiajun Chen}.} \bibinfo{year}{2021}\natexlab{}.
\newblock \showarticletitle{Dual side deep context-aware modulation for social recommendation}. In \bibinfo{booktitle}{\emph{WWW}}. \bibinfo{pages}{2524--2534}.
\newblock


\bibitem[Gao et~al\mbox{.}(2022)]%
        {gao2022self}
\bibfield{author}{\bibinfo{person}{Yunjun Gao}, \bibinfo{person}{Yuntao Du}, \bibinfo{person}{Yujia Hu}, \bibinfo{person}{Lu Chen}, \bibinfo{person}{Xinjun Zhu}, \bibinfo{person}{Ziquan Fang}, {and} \bibinfo{person}{Baihua Zheng}.} \bibinfo{year}{2022}\natexlab{}.
\newblock \showarticletitle{Self-guided learning to denoise for robust recommendation}. In \bibinfo{booktitle}{\emph{SIGIR}}. \bibinfo{pages}{1412--1422}.
\newblock


\bibitem[Gretton et~al\mbox{.}(2005)]%
        {gretton2005measuring}
\bibfield{author}{\bibinfo{person}{Arthur Gretton}, \bibinfo{person}{Olivier Bousquet}, \bibinfo{person}{Alex Smola}, {and} \bibinfo{person}{Bernhard Sch{\"o}lkopf}.} \bibinfo{year}{2005}\natexlab{}.
\newblock \showarticletitle{Measuring statistical dependence with Hilbert-Schmidt norms}. In \bibinfo{booktitle}{\emph{International conference on algorithmic learning theory}}. Springer, \bibinfo{pages}{63--77}.
\newblock


\bibitem[Gronowski et~al\mbox{.}(2023)]%
        {gronowski2023classification}
\bibfield{author}{\bibinfo{person}{Adam Gronowski}, \bibinfo{person}{William Paul}, \bibinfo{person}{Fady Alajaji}, \bibinfo{person}{Bahman Gharesifard}, {and} \bibinfo{person}{Philippe Burlina}.} \bibinfo{year}{2023}\natexlab{}.
\newblock \showarticletitle{Classification utility, fairness, and compactness via tunable information bottleneck and R{\'e}nyi measures}.
\newblock \bibinfo{journal}{\emph{IEEE Transactions on Information Forensics and Security}} (\bibinfo{year}{2023}).
\newblock


\bibitem[Gunawardana and Shani(2009)]%
        {gunawardana2009survey}
\bibfield{author}{\bibinfo{person}{Asela Gunawardana} {and} \bibinfo{person}{Guy Shani}.} \bibinfo{year}{2009}\natexlab{}.
\newblock \showarticletitle{A survey of accuracy evaluation metrics of recommendation tasks.}
\newblock \bibinfo{journal}{\emph{Journal of Machine Learning Research}} \bibinfo{volume}{10}, \bibinfo{number}{12} (\bibinfo{year}{2009}).
\newblock


\bibitem[Guo et~al\mbox{.}(2015)]%
        {guo2015trustsvd}
\bibfield{author}{\bibinfo{person}{Guibing Guo}, \bibinfo{person}{Jie Zhang}, {and} \bibinfo{person}{Neil Yorke-Smith}.} \bibinfo{year}{2015}\natexlab{}.
\newblock \showarticletitle{Trustsvd: Collaborative filtering with both the explicit and implicit influence of user trust and of item ratings}. In \bibinfo{booktitle}{\emph{AAAI}}, Vol.~\bibinfo{volume}{29}.
\newblock


\bibitem[He et~al\mbox{.}(2020)]%
        {LightGCN}
\bibfield{author}{\bibinfo{person}{Xiangnan He}, \bibinfo{person}{Kuan Deng}, \bibinfo{person}{Xiang Wang}, \bibinfo{person}{Yan Li}, \bibinfo{person}{Yongdong Zhang}, {and} \bibinfo{person}{Meng Wang}.} \bibinfo{year}{2020}\natexlab{}.
\newblock \showarticletitle{Lightgcn: Simplifying and powering graph convolution network for recommendation}. In \bibinfo{booktitle}{\emph{SIGIR}}. \bibinfo{pages}{639--648}.
\newblock


\bibitem[He et~al\mbox{.}(2024)]%
        {he2024double}
\bibfield{author}{\bibinfo{person}{Zhuangzhuang He}, \bibinfo{person}{Yifan Wang}, \bibinfo{person}{Yonghui Yang}, \bibinfo{person}{Peijie Sun}, \bibinfo{person}{Le Wu}, \bibinfo{person}{Haoyue Bai}, \bibinfo{person}{Jinqi Gong}, \bibinfo{person}{Richang Hong}, {and} \bibinfo{person}{Min Zhang}.} \bibinfo{year}{2024}\natexlab{}.
\newblock \showarticletitle{Double Correction Framework for Denoising Recommendation}.
\newblock \bibinfo{journal}{\emph{arXiv preprint arXiv:2405.11272}} (\bibinfo{year}{2024}).
\newblock


\bibitem[Hled{\'\i}k et~al\mbox{.}(2019)]%
        {hledik2019tight}
\bibfield{author}{\bibinfo{person}{Michal Hled{\'\i}k}, \bibinfo{person}{Thomas~R Sokolowski}, {and} \bibinfo{person}{Ga{\v{s}}per Tka{\v{c}}ik}.} \bibinfo{year}{2019}\natexlab{}.
\newblock \showarticletitle{A tight upper bound on mutual information}. In \bibinfo{booktitle}{\emph{2019 IEEE ITW}}. IEEE, \bibinfo{pages}{1--5}.
\newblock


\bibitem[Jamali and Ester(2010)]%
        {jamali2010matrix}
\bibfield{author}{\bibinfo{person}{Mohsen Jamali} {and} \bibinfo{person}{Martin Ester}.} \bibinfo{year}{2010}\natexlab{}.
\newblock \showarticletitle{A matrix factorization technique with trust propagation for recommendation in social networks}. In \bibinfo{booktitle}{\emph{Recsys}}. \bibinfo{pages}{135--142}.
\newblock


\bibitem[Jang et~al\mbox{.}(2017)]%
        {jang2016categorical}
\bibfield{author}{\bibinfo{person}{Eric Jang}, \bibinfo{person}{Shixiang Gu}, {and} \bibinfo{person}{Ben Poole}.} \bibinfo{year}{2017}\natexlab{}.
\newblock \showarticletitle{Categorical reparameterization with gumbel-softmax}.
\newblock \bibinfo{journal}{\emph{ICLR}} (\bibinfo{year}{2017}).
\newblock


\bibitem[Jiang et~al\mbox{.}(2014)]%
        {jiang2014scalable}
\bibfield{author}{\bibinfo{person}{Meng Jiang}, \bibinfo{person}{Peng Cui}, \bibinfo{person}{Fei Wang}, \bibinfo{person}{Wenwu Zhu}, {and} \bibinfo{person}{Shiqiang Yang}.} \bibinfo{year}{2014}\natexlab{}.
\newblock \showarticletitle{Scalable recommendation with social contextual information}.
\newblock \bibinfo{journal}{\emph{IEEE TKDE}} \bibinfo{volume}{26}, \bibinfo{number}{11} (\bibinfo{year}{2014}), \bibinfo{pages}{2789--2802}.
\newblock


\bibitem[Kipf and Welling(2017)]%
        {ICLR2017GCN}
\bibfield{author}{\bibinfo{person}{Thomas~N Kipf} {and} \bibinfo{person}{Max Welling}.} \bibinfo{year}{2017}\natexlab{}.
\newblock \showarticletitle{Semi-supervised classification with graph convolutional networks}. In \bibinfo{booktitle}{\emph{ICLR}}.
\newblock


\bibitem[Konstas et~al\mbox{.}(2009)]%
        {konstas2009social}
\bibfield{author}{\bibinfo{person}{Ioannis Konstas}, \bibinfo{person}{Vassilios Stathopoulos}, {and} \bibinfo{person}{Joemon~M Jose}.} \bibinfo{year}{2009}\natexlab{}.
\newblock \showarticletitle{On social networks and collaborative recommendation}. In \bibinfo{booktitle}{\emph{SIGIR}}. \bibinfo{pages}{195--202}.
\newblock


\bibitem[Krichene and Rendle(2020)]%
        {krichene2020sampled}
\bibfield{author}{\bibinfo{person}{Walid Krichene} {and} \bibinfo{person}{Steffen Rendle}.} \bibinfo{year}{2020}\natexlab{}.
\newblock \showarticletitle{On sampled metrics for item recommendation}. In \bibinfo{booktitle}{\emph{SIGKDD}}. \bibinfo{pages}{1748--1757}.
\newblock


\bibitem[Liao et~al\mbox{.}(2022)]%
        {liao2022sociallgn}
\bibfield{author}{\bibinfo{person}{Jie Liao}, \bibinfo{person}{Wei Zhou}, \bibinfo{person}{Fengji Luo}, \bibinfo{person}{Junhao Wen}, \bibinfo{person}{Min Gao}, \bibinfo{person}{Xiuhua Li}, {and} \bibinfo{person}{Jun Zeng}.} \bibinfo{year}{2022}\natexlab{}.
\newblock \showarticletitle{SocialLGN: Light graph convolution network for social recommendation}.
\newblock \bibinfo{journal}{\emph{Information Sciences}}  \bibinfo{volume}{589} (\bibinfo{year}{2022}), \bibinfo{pages}{595--607}.
\newblock


\bibitem[Ma et~al\mbox{.}(2008)]%
        {ma2008sorec}
\bibfield{author}{\bibinfo{person}{Hao Ma}, \bibinfo{person}{Haixuan Yang}, \bibinfo{person}{Michael~R Lyu}, {and} \bibinfo{person}{Irwin King}.} \bibinfo{year}{2008}\natexlab{}.
\newblock \showarticletitle{Sorec: social recommendation using probabilistic matrix factorization}. In \bibinfo{booktitle}{\emph{CIKM}}. \bibinfo{pages}{931--940}.
\newblock


\bibitem[Ma et~al\mbox{.}(2011)]%
        {ma2011recommender}
\bibfield{author}{\bibinfo{person}{Hao Ma}, \bibinfo{person}{Dengyong Zhou}, \bibinfo{person}{Chao Liu}, \bibinfo{person}{Michael~R Lyu}, {and} \bibinfo{person}{Irwin King}.} \bibinfo{year}{2011}\natexlab{}.
\newblock \showarticletitle{Recommender systems with social regularization}. In \bibinfo{booktitle}{\emph{WSDM}}. \bibinfo{pages}{287--296}.
\newblock


\bibitem[Ma et~al\mbox{.}(2020)]%
        {ma2020HSIC}
\bibfield{author}{\bibinfo{person}{Wan-Duo~Kurt Ma}, \bibinfo{person}{JP Lewis}, {and} \bibinfo{person}{W~Bastiaan Kleijn}.} \bibinfo{year}{2020}\natexlab{}.
\newblock \showarticletitle{The HSIC bottleneck: Deep learning without back-propagation}. In \bibinfo{booktitle}{\emph{AAAI}}, Vol.~\bibinfo{volume}{34}. \bibinfo{pages}{5085--5092}.
\newblock


\bibitem[Marsden and Friedkin(1993)]%
        {marsden1993network}
\bibfield{author}{\bibinfo{person}{Peter~V Marsden} {and} \bibinfo{person}{Noah~E Friedkin}.} \bibinfo{year}{1993}\natexlab{}.
\newblock \showarticletitle{Network studies of social influence}.
\newblock \bibinfo{journal}{\emph{Sociological Methods \& Research}} \bibinfo{volume}{22}, \bibinfo{number}{1} (\bibinfo{year}{1993}), \bibinfo{pages}{127--151}.
\newblock


\bibitem[McPherson et~al\mbox{.}(2001)]%
        {mcpherson2001birds}
\bibfield{author}{\bibinfo{person}{Miller McPherson}, \bibinfo{person}{Lynn Smith-Lovin}, {and} \bibinfo{person}{James~M Cook}.} \bibinfo{year}{2001}\natexlab{}.
\newblock \showarticletitle{Birds of a feather: Homophily in social networks}.
\newblock \bibinfo{journal}{\emph{Annual review of sociology}} \bibinfo{volume}{27}, \bibinfo{number}{1} (\bibinfo{year}{2001}), \bibinfo{pages}{415--444}.
\newblock


\bibitem[Mnih and Salakhutdinov(2007)]%
        {PMF}
\bibfield{author}{\bibinfo{person}{Andriy Mnih} {and} \bibinfo{person}{Russ~R Salakhutdinov}.} \bibinfo{year}{2007}\natexlab{}.
\newblock \showarticletitle{Probabilistic matrix factorization}.
\newblock \bibinfo{journal}{\emph{NeurIPS}}  \bibinfo{volume}{20} (\bibinfo{year}{2007}).
\newblock


\bibitem[Oord et~al\mbox{.}(2018)]%
        {oord2018InfoNCE}
\bibfield{author}{\bibinfo{person}{Aaron van~den Oord}, \bibinfo{person}{Yazhe Li}, {and} \bibinfo{person}{Oriol Vinyals}.} \bibinfo{year}{2018}\natexlab{}.
\newblock \showarticletitle{Representation learning with contrastive predictive coding}.
\newblock \bibinfo{journal}{\emph{arXiv preprint arXiv:1807.03748}} (\bibinfo{year}{2018}).
\newblock


\bibitem[Pan et~al\mbox{.}(2020)]%
        {pan2020correlative}
\bibfield{author}{\bibinfo{person}{Yiteng Pan}, \bibinfo{person}{Fazhi He}, {and} \bibinfo{person}{Haiping Yu}.} \bibinfo{year}{2020}\natexlab{}.
\newblock \showarticletitle{A correlative denoising autoencoder to model social influence for top-N recommender system}.
\newblock \bibinfo{journal}{\emph{Frontiers of Computer science}}  \bibinfo{volume}{14} (\bibinfo{year}{2020}), \bibinfo{pages}{1--13}.
\newblock


\bibitem[Patrick et~al\mbox{.}(2021)]%
        {patrick2020support}
\bibfield{author}{\bibinfo{person}{Mandela Patrick}, \bibinfo{person}{Po-Yao Huang}, \bibinfo{person}{Yuki Asano}, \bibinfo{person}{Florian Metze}, \bibinfo{person}{Alexander Hauptmann}, \bibinfo{person}{Joao Henriques}, {and} \bibinfo{person}{Andrea Vedaldi}.} \bibinfo{year}{2021}\natexlab{}.
\newblock \showarticletitle{Support-set bottlenecks for video-text representation learning}.
\newblock \bibinfo{journal}{\emph{ICLR}} (\bibinfo{year}{2021}).
\newblock


\bibitem[Quan et~al\mbox{.}(2023)]%
        {WWW2023robust}
\bibfield{author}{\bibinfo{person}{Yuhan Quan}, \bibinfo{person}{Jingtao Ding}, \bibinfo{person}{Chen Gao}, \bibinfo{person}{Lingling Yi}, \bibinfo{person}{Depeng Jin}, {and} \bibinfo{person}{Yong Li}.} \bibinfo{year}{2023}\natexlab{}.
\newblock \showarticletitle{Robust Preference-Guided Denoising for Graph based Social Recommendation}. In \bibinfo{booktitle}{\emph{WWW}}. \bibinfo{pages}{1097--1108}.
\newblock


\bibitem[Rendle et~al\mbox{.}(2009)]%
        {UAI2009BPR}
\bibfield{author}{\bibinfo{person}{Steffen Rendle}, \bibinfo{person}{Christoph Freudenthaler}, \bibinfo{person}{Zeno Gantner}, {and} \bibinfo{person}{Lars Schmidt-Thieme}.} \bibinfo{year}{2009}\natexlab{}.
\newblock \showarticletitle{BPR: Bayesian personalized ranking from implicit feedback}. In \bibinfo{booktitle}{\emph{UAI}}. \bibinfo{pages}{452--461}.
\newblock


\bibitem[Saxe et~al\mbox{.}(2019)]%
        {saxe2019information}
\bibfield{author}{\bibinfo{person}{Andrew~M Saxe}, \bibinfo{person}{Yamini Bansal}, \bibinfo{person}{Joel Dapello}, \bibinfo{person}{Madhu Advani}, \bibinfo{person}{Artemy Kolchinsky}, \bibinfo{person}{Brendan~D Tracey}, {and} \bibinfo{person}{David~D Cox}.} \bibinfo{year}{2019}\natexlab{}.
\newblock \showarticletitle{On the information bottleneck theory of deep learning}.
\newblock \bibinfo{journal}{\emph{Journal of Statistical Mechanics: Theory and Experiment}} \bibinfo{volume}{2019}, \bibinfo{number}{12} (\bibinfo{year}{2019}), \bibinfo{pages}{124020}.
\newblock


\bibitem[Shao et~al\mbox{.}(2022)]%
        {shao2022faircf}
\bibfield{author}{\bibinfo{person}{Pengyang Shao}, \bibinfo{person}{Le Wu}, \bibinfo{person}{Lei Chen}, \bibinfo{person}{Kun Zhang}, {and} \bibinfo{person}{Meng Wang}.} \bibinfo{year}{2022}\natexlab{}.
\newblock \showarticletitle{FairCF: Fairness-aware collaborative filtering}.
\newblock \bibinfo{journal}{\emph{Science China Information Sciences}} \bibinfo{volume}{65}, \bibinfo{number}{12} (\bibinfo{year}{2022}), \bibinfo{pages}{222102}.
\newblock


\bibitem[Shao et~al\mbox{.}(2024)]%
        {shao2024average}
\bibfield{author}{\bibinfo{person}{Pengyang Shao}, \bibinfo{person}{Le Wu}, \bibinfo{person}{Kun Zhang}, \bibinfo{person}{Defu Lian}, \bibinfo{person}{Richang Hong}, \bibinfo{person}{Yong Li}, {and} \bibinfo{person}{Meng Wang}.} \bibinfo{year}{2024}\natexlab{}.
\newblock \showarticletitle{Average User-side Counterfactual Fairness for Collaborative Filtering}.
\newblock \bibinfo{journal}{\emph{ACM TOIS}} (\bibinfo{year}{2024}).
\newblock


\bibitem[Shapiro(2003)]%
        {shapiro2003monte}
\bibfield{author}{\bibinfo{person}{Alexander Shapiro}.} \bibinfo{year}{2003}\natexlab{}.
\newblock \showarticletitle{Monte Carlo sampling methods}.
\newblock \bibinfo{journal}{\emph{Handbooks in operations research and management science}}  \bibinfo{volume}{10} (\bibinfo{year}{2003}), \bibinfo{pages}{353--425}.
\newblock


\bibitem[Steck(2013)]%
        {steck2013evaluation}
\bibfield{author}{\bibinfo{person}{Harald Steck}.} \bibinfo{year}{2013}\natexlab{}.
\newblock \showarticletitle{Evaluation of recommendations: rating-prediction and ranking}. In \bibinfo{booktitle}{\emph{RecSys}}. \bibinfo{pages}{213--220}.
\newblock


\bibitem[Sun et~al\mbox{.}(2018)]%
        {sun2018attentive}
\bibfield{author}{\bibinfo{person}{Peijie Sun}, \bibinfo{person}{Le Wu}, {and} \bibinfo{person}{Meng Wang}.} \bibinfo{year}{2018}\natexlab{}.
\newblock \showarticletitle{Attentive recurrent social recommendation}. In \bibinfo{booktitle}{\emph{SIGIR}}. \bibinfo{pages}{185--194}.
\newblock


\bibitem[Sun(2023)]%
        {sun2023denoising}
\bibfield{author}{\bibinfo{person}{Youchen Sun}.} \bibinfo{year}{2023}\natexlab{}.
\newblock \showarticletitle{Denoising Explicit Social Signals for Robust Recommendation}. In \bibinfo{booktitle}{\emph{RecSys}}. \bibinfo{pages}{1344--1348}.
\newblock


\bibitem[Tang et~al\mbox{.}(2013)]%
        {tang2013social}
\bibfield{author}{\bibinfo{person}{Jiliang Tang}, \bibinfo{person}{Xia Hu}, {and} \bibinfo{person}{Huan Liu}.} \bibinfo{year}{2013}\natexlab{}.
\newblock \showarticletitle{Social recommendation: a review}.
\newblock \bibinfo{journal}{\emph{Social Network Analysis and Mining}}  \bibinfo{volume}{3} (\bibinfo{year}{2013}), \bibinfo{pages}{1113--1133}.
\newblock


\bibitem[Tishby et~al\mbox{.}(2000)]%
        {tishby2000information}
\bibfield{author}{\bibinfo{person}{Naftali Tishby}, \bibinfo{person}{Fernando~C Pereira}, {and} \bibinfo{person}{William Bialek}.} \bibinfo{year}{2000}\natexlab{}.
\newblock \showarticletitle{The information bottleneck method}.
\newblock \bibinfo{journal}{\emph{arXiv preprint physics/0004057}} (\bibinfo{year}{2000}).
\newblock


\bibitem[Tishby and Zaslavsky(2015)]%
        {tishby2015deep}
\bibfield{author}{\bibinfo{person}{Naftali Tishby} {and} \bibinfo{person}{Noga Zaslavsky}.} \bibinfo{year}{2015}\natexlab{}.
\newblock \showarticletitle{Deep learning and the information bottleneck principle}. In \bibinfo{booktitle}{\emph{2015 IEEE ITW}}. IEEE, \bibinfo{pages}{1--5}.
\newblock


\bibitem[Veli{\v{c}}kovi{\'c} et~al\mbox{.}(2018)]%
        {ICLR2018GAT}
\bibfield{author}{\bibinfo{person}{Petar Veli{\v{c}}kovi{\'c}}, \bibinfo{person}{Guillem Cucurull}, \bibinfo{person}{Arantxa Casanova}, \bibinfo{person}{Adriana Romero}, \bibinfo{person}{Pietro Lio}, {and} \bibinfo{person}{Yoshua Bengio}.} \bibinfo{year}{2018}\natexlab{}.
\newblock \showarticletitle{Graph attention networks}. In \bibinfo{booktitle}{\emph{ICLR}}.
\newblock


\bibitem[Vert et~al\mbox{.}(2004)]%
        {vert2004kernel}
\bibfield{author}{\bibinfo{person}{Jean-Philippe Vert}, \bibinfo{person}{Koji Tsuda}, {and} \bibinfo{person}{Bernhard Sch{\"o}lkopf}.} \bibinfo{year}{2004}\natexlab{}.
\newblock \showarticletitle{A primer on kernel methods}.
\newblock \bibinfo{journal}{\emph{Kernel methods in computational biology}}  \bibinfo{volume}{47} (\bibinfo{year}{2004}), \bibinfo{pages}{35--70}.
\newblock


\bibitem[Wang et~al\mbox{.}(2023b)]%
        {wang2023learning}
\bibfield{author}{\bibinfo{person}{Bowen Wang}, \bibinfo{person}{Liangzhi Li}, \bibinfo{person}{Yuta Nakashima}, {and} \bibinfo{person}{Hajime Nagahara}.} \bibinfo{year}{2023}\natexlab{b}.
\newblock \showarticletitle{Learning Bottleneck Concepts in Image Classification}. In \bibinfo{booktitle}{\emph{CVPR}}. \bibinfo{pages}{10962--10971}.
\newblock


\bibitem[Wang et~al\mbox{.}(2021c)]%
        {TOIS2021hypersorec}
\bibfield{author}{\bibinfo{person}{Hao Wang}, \bibinfo{person}{Defu Lian}, \bibinfo{person}{Hanghang Tong}, \bibinfo{person}{Qi Liu}, \bibinfo{person}{Zhenya Huang}, {and} \bibinfo{person}{Enhong Chen}.} \bibinfo{year}{2021}\natexlab{c}.
\newblock \showarticletitle{Hypersorec: Exploiting hyperbolic user and item representations with multiple aspects for social-aware recommendation}.
\newblock \bibinfo{journal}{\emph{TOIS}} \bibinfo{volume}{40}, \bibinfo{number}{2} (\bibinfo{year}{2021}), \bibinfo{pages}{1--28}.
\newblock


\bibitem[Wang et~al\mbox{.}(2021a)]%
        {wang2021denoising}
\bibfield{author}{\bibinfo{person}{Wenjie Wang}, \bibinfo{person}{Fuli Feng}, \bibinfo{person}{Xiangnan He}, \bibinfo{person}{Liqiang Nie}, {and} \bibinfo{person}{Tat-Seng Chua}.} \bibinfo{year}{2021}\natexlab{a}.
\newblock \showarticletitle{Denoising implicit feedback for recommendation}. In \bibinfo{booktitle}{\emph{WSDM}}. \bibinfo{pages}{373--381}.
\newblock


\bibitem[Wang et~al\mbox{.}(2023a)]%
        {wang2023efficient}
\bibfield{author}{\bibinfo{person}{Zongwei Wang}, \bibinfo{person}{Min Gao}, \bibinfo{person}{Wentao Li}, \bibinfo{person}{Junliang Yu}, \bibinfo{person}{Linxin Guo}, {and} \bibinfo{person}{Hongzhi Yin}.} \bibinfo{year}{2023}\natexlab{a}.
\newblock \showarticletitle{Efficient bi-level optimization for recommendation denoising}. In \bibinfo{booktitle}{\emph{SIGKDD}}. \bibinfo{pages}{2502--2511}.
\newblock


\bibitem[Wang et~al\mbox{.}(2021b)]%
        {wang2021revisiting}
\bibfield{author}{\bibinfo{person}{Zifeng Wang}, \bibinfo{person}{Tong Jian}, \bibinfo{person}{Aria Masoomi}, \bibinfo{person}{Stratis Ioannidis}, {and} \bibinfo{person}{Jennifer Dy}.} \bibinfo{year}{2021}\natexlab{b}.
\newblock \showarticletitle{Revisiting hilbert-schmidt information bottleneck for adversarial robustness}.
\newblock \bibinfo{journal}{\emph{NeurIPS}}  \bibinfo{volume}{34} (\bibinfo{year}{2021}), \bibinfo{pages}{586--597}.
\newblock


\bibitem[Wu et~al\mbox{.}(2020a)]%
        {wu2020diffnet++}
\bibfield{author}{\bibinfo{person}{Le Wu}, \bibinfo{person}{Junwei Li}, \bibinfo{person}{Peijie Sun}, \bibinfo{person}{Richang Hong}, \bibinfo{person}{Yong Ge}, {and} \bibinfo{person}{Meng Wang}.} \bibinfo{year}{2020}\natexlab{a}.
\newblock \showarticletitle{Diffnet++: A neural influence and interest diffusion network for social recommendation}.
\newblock \bibinfo{journal}{\emph{TKDE}} (\bibinfo{year}{2020}).
\newblock


\bibitem[Wu et~al\mbox{.}(2019)]%
        {DiffNet}
\bibfield{author}{\bibinfo{person}{Le Wu}, \bibinfo{person}{Peijie Sun}, \bibinfo{person}{Yanjie Fu}, \bibinfo{person}{Richang Hong}, \bibinfo{person}{Xiting Wang}, {and} \bibinfo{person}{Meng Wang}.} \bibinfo{year}{2019}\natexlab{}.
\newblock \showarticletitle{A neural influence diffusion model for social recommendation}. In \bibinfo{booktitle}{\emph{SIGIR}}. \bibinfo{pages}{235--244}.
\newblock


\bibitem[Wu et~al\mbox{.}(2020d)]%
        {wu2020learning}
\bibfield{author}{\bibinfo{person}{Le Wu}, \bibinfo{person}{Yonghui Yang}, \bibinfo{person}{Lei Chen}, \bibinfo{person}{Defu Lian}, \bibinfo{person}{Richang Hong}, {and} \bibinfo{person}{Meng Wang}.} \bibinfo{year}{2020}\natexlab{d}.
\newblock \showarticletitle{Learning to transfer graph embeddings for inductive graph based recommendation}. In \bibinfo{booktitle}{\emph{SIGIR}}. \bibinfo{pages}{1211--1220}.
\newblock


\bibitem[Wu et~al\mbox{.}(2020e)]%
        {wu2020joint}
\bibfield{author}{\bibinfo{person}{Le Wu}, \bibinfo{person}{Yonghui Yang}, \bibinfo{person}{Kun Zhang}, \bibinfo{person}{Richang Hong}, \bibinfo{person}{Yanjie Fu}, {and} \bibinfo{person}{Meng Wang}.} \bibinfo{year}{2020}\natexlab{e}.
\newblock \showarticletitle{Joint item recommendation and attribute inference: An adaptive graph convolutional network approach}. In \bibinfo{booktitle}{\emph{SIGIR}}. \bibinfo{pages}{679--688}.
\newblock


\bibitem[Wu et~al\mbox{.}(2022)]%
        {wu2022graph}
\bibfield{author}{\bibinfo{person}{Shiwen Wu}, \bibinfo{person}{Fei Sun}, \bibinfo{person}{Wentao Zhang}, \bibinfo{person}{Xu Xie}, {and} \bibinfo{person}{Bin Cui}.} \bibinfo{year}{2022}\natexlab{}.
\newblock \showarticletitle{Graph neural networks in recommender systems: a survey}.
\newblock \bibinfo{journal}{\emph{Comput. Surveys}} \bibinfo{volume}{55}, \bibinfo{number}{5} (\bibinfo{year}{2022}), \bibinfo{pages}{1--37}.
\newblock


\bibitem[Wu et~al\mbox{.}(2020c)]%
        {wu2020GIB}
\bibfield{author}{\bibinfo{person}{Tailin Wu}, \bibinfo{person}{Hongyu Ren}, \bibinfo{person}{Pan Li}, {and} \bibinfo{person}{Jure Leskovec}.} \bibinfo{year}{2020}\natexlab{c}.
\newblock \showarticletitle{Graph information bottleneck}.
\newblock \bibinfo{journal}{\emph{NeurIPS}}  \bibinfo{volume}{33} (\bibinfo{year}{2020}), \bibinfo{pages}{20437--20448}.
\newblock


\bibitem[Wu et~al\mbox{.}(2020b)]%
        {wu2020comprehensive}
\bibfield{author}{\bibinfo{person}{Zonghan Wu}, \bibinfo{person}{Shirui Pan}, \bibinfo{person}{Fengwen Chen}, \bibinfo{person}{Guodong Long}, \bibinfo{person}{Chengqi Zhang}, {and} \bibinfo{person}{S~Yu Philip}.} \bibinfo{year}{2020}\natexlab{b}.
\newblock \showarticletitle{A comprehensive survey on graph neural networks}.
\newblock \bibinfo{journal}{\emph{IEEE TNNLS}} \bibinfo{volume}{32}, \bibinfo{number}{1} (\bibinfo{year}{2020}), \bibinfo{pages}{4--24}.
\newblock


\bibitem[Yang et~al\mbox{.}(2021)]%
        {yang2021enhanced}
\bibfield{author}{\bibinfo{person}{Yonghui Yang}, \bibinfo{person}{Le Wu}, \bibinfo{person}{Richang Hong}, \bibinfo{person}{Kun Zhang}, {and} \bibinfo{person}{Meng Wang}.} \bibinfo{year}{2021}\natexlab{}.
\newblock \showarticletitle{Enhanced graph learning for collaborative filtering via mutual information maximization}. In \bibinfo{booktitle}{\emph{SIGIR}}. \bibinfo{pages}{71--80}.
\newblock


\bibitem[Yang et~al\mbox{.}(2023b)]%
        {HGSR}
\bibfield{author}{\bibinfo{person}{Yonghui Yang}, \bibinfo{person}{Le Wu}, \bibinfo{person}{Kun Zhang}, \bibinfo{person}{Richang Hong}, \bibinfo{person}{Hailin Zhou}, \bibinfo{person}{Zhiqiang Zhang}, \bibinfo{person}{Jun Zhou}, {and} \bibinfo{person}{Meng Wang}.} \bibinfo{year}{2023}\natexlab{b}.
\newblock \showarticletitle{Hyperbolic Graph Learning for Social Recommendation}.
\newblock \bibinfo{journal}{\emph{IEEE TKDE}} (\bibinfo{year}{2023}).
\newblock


\bibitem[Yang et~al\mbox{.}(2023a)]%
        {yang2023generative}
\bibfield{author}{\bibinfo{person}{Yonghui Yang}, \bibinfo{person}{Zhengwei Wu}, \bibinfo{person}{Le Wu}, \bibinfo{person}{Kun Zhang}, \bibinfo{person}{Richang Hong}, \bibinfo{person}{Zhiqiang Zhang}, \bibinfo{person}{Jun Zhou}, {and} \bibinfo{person}{Meng Wang}.} \bibinfo{year}{2023}\natexlab{a}.
\newblock \showarticletitle{Generative-contrastive graph learning for recommendation}. In \bibinfo{booktitle}{\emph{SIGIR}}. \bibinfo{pages}{1117--1126}.
\newblock


\bibitem[Yu et~al\mbox{.}(2019)]%
        {yu2019generating}
\bibfield{author}{\bibinfo{person}{Junliang Yu}, \bibinfo{person}{Min Gao}, \bibinfo{person}{Hongzhi Yin}, \bibinfo{person}{Jundong Li}, \bibinfo{person}{Chongming Gao}, {and} \bibinfo{person}{Qinyong Wang}.} \bibinfo{year}{2019}\natexlab{}.
\newblock \showarticletitle{Generating reliable friends via adversarial training to improve social recommendation}. In \bibinfo{booktitle}{\emph{ICDE}}. IEEE, \bibinfo{pages}{768--777}.
\newblock


\bibitem[Yu et~al\mbox{.}(2021a)]%
        {SEPT}
\bibfield{author}{\bibinfo{person}{Junliang Yu}, \bibinfo{person}{Hongzhi Yin}, \bibinfo{person}{Min Gao}, \bibinfo{person}{Xin Xia}, \bibinfo{person}{Xiangliang Zhang}, {and} \bibinfo{person}{Nguyen~Quoc Viet~Hung}.} \bibinfo{year}{2021}\natexlab{a}.
\newblock \showarticletitle{Socially-aware self-supervised tri-training for recommendation}. In \bibinfo{booktitle}{\emph{SIGKDD}}. \bibinfo{pages}{2084--2092}.
\newblock


\bibitem[Yu et~al\mbox{.}(2020)]%
        {TKDE2020enhancing}
\bibfield{author}{\bibinfo{person}{Junliang Yu}, \bibinfo{person}{Hongzhi Yin}, \bibinfo{person}{Jundong Li}, \bibinfo{person}{Min Gao}, \bibinfo{person}{Zi Huang}, {and} \bibinfo{person}{Lizhen Cui}.} \bibinfo{year}{2020}\natexlab{}.
\newblock \showarticletitle{Enhancing social recommendation with adversarial graph convolutional networks}.
\newblock \bibinfo{journal}{\emph{IEEE TKDE}} \bibinfo{volume}{34}, \bibinfo{number}{8} (\bibinfo{year}{2020}), \bibinfo{pages}{3727--3739}.
\newblock


\bibitem[Yu et~al\mbox{.}(2021b)]%
        {yu2021self}
\bibfield{author}{\bibinfo{person}{Junliang Yu}, \bibinfo{person}{Hongzhi Yin}, \bibinfo{person}{Jundong Li}, \bibinfo{person}{Qinyong Wang}, \bibinfo{person}{Nguyen Quoc~Viet Hung}, {and} \bibinfo{person}{Xiangliang Zhang}.} \bibinfo{year}{2021}\natexlab{b}.
\newblock \showarticletitle{Self-supervised multi-channel hypergraph convolutional network for social recommendation}. In \bibinfo{booktitle}{\emph{WWW}}. \bibinfo{pages}{413--424}.
\newblock


\bibitem[Zhao et~al\mbox{.}(2020)]%
        {zhao2020revisiting}
\bibfield{author}{\bibinfo{person}{Wayne~Xin Zhao}, \bibinfo{person}{Junhua Chen}, \bibinfo{person}{Pengfei Wang}, \bibinfo{person}{Qi Gu}, {and} \bibinfo{person}{Ji-Rong Wen}.} \bibinfo{year}{2020}\natexlab{}.
\newblock \showarticletitle{Revisiting alternative experimental settings for evaluating top-n item recommendation algorithms}. In \bibinfo{booktitle}{\emph{CIKM}}. \bibinfo{pages}{2329--2332}.
\newblock


\bibitem[Zheng et~al\mbox{.}(2020)]%
        {ICML2020robust}
\bibfield{author}{\bibinfo{person}{Cheng Zheng}, \bibinfo{person}{Bo Zong}, \bibinfo{person}{Wei Cheng}, \bibinfo{person}{Dongjin Song}, \bibinfo{person}{Jingchao Ni}, \bibinfo{person}{Wenchao Yu}, \bibinfo{person}{Haifeng Chen}, {and} \bibinfo{person}{Wei Wang}.} \bibinfo{year}{2020}\natexlab{}.
\newblock \showarticletitle{Robust graph representation learning via neural sparsification}. In \bibinfo{booktitle}{\emph{ICML}}. PMLR, \bibinfo{pages}{11458--11468}.
\newblock


\end{thebibliography}
